\documentclass[paper,11pt]{JHEP}
\usepackage[centertags]{amsmath}
\usepackage{amsfonts} \usepackage{amssymb} \usepackage{amsthm}
\allowdisplaybreaks[1]
\usepackage{graphicx}
\newcommand{\Z}{{\mathbb Z}}

\newcommand{\ii}{\mathrm{i}}
\newcommand{\ee}{\mathrm{e}}

\newcommand{\cN}{{\mathcal{N}}}

\newcommand{\cS}{{\mathcal{S}}}

\newcommand{\one}{{\rm 1\kern -.9mm l}}
\newcommand{\bone}{\mathbf{1}}
\newcommand{\ft}[2]{{\textstyle\frac{#1}{#2}}}

\def\sp#1#2{\big[^{#1}_{#2}\big]}

\newdimen\tableauside\tableauside=1.0ex
\newdimen\tableaurule\tableaurule=0.4pt
\newdimen\tableaustep
\def\phantomhrule#1{\hbox{\vbox to0pt{\hrule height\tableaurule
width#1\vss}}}
\def\phantomvrule#1{\vbox{\hbox to0pt{\vrule width\tableaurule
height#1\hss}}}
\def\sqr{\vbox{%
  \phantomhrule\tableaustep
\hbox{\phantomvrule\tableaustep\kern\tableaustep\phantomvrule\tableaustep}%
  \hbox{\vbox{\phantomhrule\tableauside}\kern-\tableaurule}}}
\def\squares#1{\hbox{\count0=#1\noindent\loop\sqr
  \advance\count0 by-1 \ifnum\count0>0\repeat}}
\def\tableau#1{\vcenter{\offinterlineskip
  \tableaustep=\tableauside\advance\tableaustep by-\tableaurule
  \kern\normallineskip\hbox
    {\kern\normallineskip\vbox
      {\gettableau#1 0 }%
     \kern\normallineskip\kern\tableaurule}%
  \kern\normallineskip\kern\tableaurule}}
\def\gettableau#1 {\ifnum#1=0\let\next=\null\else
  \squares{#1}\let\next=\gettableau\fi\next}
\newcommand{\Yfund}{\tableau{1}}
\newcommand{\Ysymm}{\tableau{2}}
\newcommand{\Yasymm}{\tableau{1 1}}

\def\XXint#1#2#3{{\setbox0=\hbox{$#1{#2#3}{\int}$}
     \vcenter{\hbox{$#2#3$}}\kern-.5\wd0}}

\title{S-duality and the prepotential in $\mathcal{N}=2^\star$ theories (II): the non-simply laced algebras
}
\author{M. Bill\'o$^1$, M. Frau$^{1}$, F. Fucito$^{2}$, A. Lerda$^{3,1}$, J.F. Morales$^{2}$
\\
\vskip 0.2cm
$^1$ Universit\`a di Torino, Dipartimento di Fisica
\\ and I.N.F.N. - sezione di Torino,
Via P. Giuria 1, I-10125 Torino, Italy\\
\vskip 0.2cm
$^2$ I.N.F.N - sezione di Roma 2\\
and Universit\`a di Roma Tor Vergata, Dipartimento di Fisica\\
Via della Ricerca Scientifica, I-00133 Roma, Italy
\vskip 0.2cm
$^3$Universit\`a del Piemonte Orientale, Dipartimento di Scienze e Innovazione Tecnologica, \\
and I.N.F.N. - Gruppo Collegato di Alessandria - sezione di Torino\\
Viale T. Michel  11, I-15121 Alessandria, Italy\\
\vspace{0.35cm}
\email{billo,frau,lerda@to.infn.it; fucito,morales@roma2.infn.it}
}
\abstract{We derive a modular anomaly equation satisfied by the prepotential
of the $\mathcal{N}=2^\star$ supersymmetric theories with non-simply laced gauge algebras,
including the classical $B_r$ and $C_r$ infinite series and the exceptional $F_4$ and $G_2$ cases. 
This equation determines the exact prepotential recursively in an expansion for small 
mass in terms of quasi-modular forms of the S-duality group. 
We also discuss the behaviour of these theories under S-duality and show that the prepotential 
of the SO$(2r+1)$ theory is mapped to that of the Sp$(2r)$ theory and viceversa,
while the exceptional $F_4$ and $G_2$ theories are mapped into themselves (up to a rotation of the roots)
in analogy with what happens for the $\mathcal{N}=4$ supersymmetric theories.
These results extend the analysis for the simply laced groups presented in a companion paper.
}


\preprint{ROM2F/2015/8}

\begin{document}
\section{Introduction}
\label{secn:intro}
In a companion paper \cite{Billo} we have studied $\cN=2^\star$ super Yang-Mills
theories with gauge groups of ADE type, generalizing and extending results that were previously
obtained for SU(2) and SU($N$) gauge groups
\cite{Huang:2011qx}\nocite{Billo:2013jba,Billo:2013fi,Huang:2013eja}\,-\,\cite{Billo:2014bja}.
The $\cN=2^\star$ theories possess eight supercharges and interpolate between the $\cN=4$ and 
the pure $\cN=2$ super Yang-Mills theories.
Their low-energy effective dynamics is encoded in the prepotential $F$ which can be conveniently 
organized as an expansion in even powers of the mass $m$ of the matter hypermultiplet.

Given that $F$ has mass-dimension two and that the only other dimensionful parameter available in the model
is the vacuum expectation value $a$ of the scalar field in the gauge vector multiplet,
each term of the prepotential at order $2n$ in the mass must be accompanied by a function of $a$ with
mass-dimension $(2-2n)$. These functions of $a$ turn out to be nicely written as sums over the root
lattice of the gauge algebra. Thus the prepotential always maintains the same form for all ADE algebras since
what changes from case to case is only the explicit expression of the roots. The coefficients in front of these
lattice sums are universal functions of the gauge coupling constant and receive both perturbative contributions at
1-loop and non-perturbative corrections due to instantons. Actually, all these contributions can be resummed
into \emph{exact} functions of the gauge coupling that are built out of the Eisenstein series, including the
second Eisenstein series $E_2$ which has an ``anomalous" behaviour under the modular transformations
of $\mathrm{Sl}(2,\mathbb{Z})$.
The presence of $E_2$ leads to a modular anomaly equation which can be put in the form of
a recursion relation for the coefficients of the mass expansion of the prepotential and encodes all
information implied by S-duality \cite{Billo} %
\footnote{Modular anomaly equations have been considered also in the context of $\cN=2$ conformal
SQCD models with fundamental matter \cite{Billo:2013fi,Billo:2013jba,Ashok:2015cba}.}.

In this paper we extend and generalize these results to $\mathcal{N}=2^\star$ theories
with non-simply laced gauge algebras $\mathfrak{g}\in \{B_r,C_r,F_4,G_2\}$. The presence of
long and short roots in these algebras
implies several important differences with respect to the ADE case, even if the overall picture remains similar.
In particular, the notion of S-duality, originally formulated 
in \cite{Montonen:1977sn}\nocite{Goddard:1976qe,Girardello:1995gf,Dorey:1996hx,Argyres:2006qr}\,-\,\cite{Kapustin:2006pk} for
the $\cN=4$ theories, can be also extended to the $\cN=2^\star$ models with non-simply laced
gauge algebras where the strong/weak coupling symmetry requirement takes the form of a
relation between the S-dual prepotential and the Legendre transform of 
its dual \cite{Seiberg:1994rs,Billo:2013jba,Billo:2013fi}. However, differently from the ADE case, 
one finds that S-duality
is not a true symmetry since it maps a theory with gauge algebra $\mathfrak{g}$
to a theory with a dual gauge algebra $\mathfrak{g}^{\!\vee}$, obtained by exchanging (and rescaling)
the long and short roots \cite{Goddard:1976qe}. This property leads to very severe constraints on the prepotential
coefficients, and it is a quite remarkable fact that they can be satisfied by imposing again
a modular anomaly equation similar to that of the ADE theories.

Another important difference with respect to the ADE case is that the modular group of the non-simply laced
theories is not $\mathrm{Sl}(2,\mathbb{Z})$ but its congruence subgroup $\Gamma_0(n_\mathfrak{g})$
\cite{Koblitz,Apostol}, where $n_{\mathfrak{g}}$ is the ratio between the norm squared of the long and short roots of $\mathfrak{g}$. Consequently, the prepotential coefficients are expressed as quasi-modular forms
of this subgroup, which include other modular functions besides the standard Eisenstein series. Such functions
as well as the Eisenstein series have a simple behaviour also under the S-duality transformation which lies outside $\Gamma_0(n_\mathfrak{g})$. Exploiting this fact together with the S-duality transformation properties of the root lattices mentioned above, it is possible to verify that the modular anomaly equation
implies the expected strong/weak coupling relation between the prepotentials of $\cN=2^\star$ theories
with dual gauge algebras.

The plan of the paper is the following: In Section~\ref{secn:sduality} we discuss the S-duality action on the gauge theories with non-simply laced algebras and derive the modular anomaly equation and the recursion relation satisfied by the quantum prepotential.
In Section~\ref{secn:multiinst} we present the microscopic computation of the instanton corrections for theories with gauge algebras in the classical $B_r$ and $C_r$ series using the equivariant localization methods
\cite{Nekrasov:2002qd}\nocite{Nekrasov:2003rj,Bruzzo:2002xf}\,-\,\cite{Fucito:2004ry}. This is necessary in order to have
explicit ``microscopic" data on the multi-instanton corrections which can then be used in order to prove S-duality or compared with the S-duality predictions. In Section~\ref{secn:recursion}
we derive exact formulas written in terms of modular forms
for the first few coefficients in the mass expansion of the prepotential for theories
with orthogonal or symplectic gauge algebras, while in Section~\ref{secn:G2F4} we repeat the analysis for the
exceptional algebras $G_2$ and $F_4$. By Taylor expanding the modular forms, one can obtain the whole series of multi-instanton corrections to the prepotential.
While for the classical algebras these can be checked against the microscopic computations, for $G_2$ and $F_4$ they are a prediction since the ADHM construction is not
known for the exceptional gauge algebras. Finally, in Section~\ref{sec:concl} we present our conclusions and perspectives. Several details about the root systems of the non-simply laced algebras and about the modular forms are contained in two technical appendices.

\section{S-duality}
\label{secn:sduality}
In this section we investigate the S-duality transformation properties of $\mathcal{N}=2^\star$ 
theories with non-simply laced gauge algebras $\mathfrak{g}\in \{B_r,C_r,F_4,G_2\}$.
In particular, we will show that the symmetry requirement
under S-duality determines the modular behaviour of the quantum prepotential and implies the 
emergence of a modular anomaly equation in the form of
a recursion relation for the coefficients of its mass expansion.

In order to treat all cases simultaneously, we introduce the convenient notation
\begin{equation}
n_{\mathfrak{g}}= \frac{\alpha_{\mathrm{L}}\cdot\alpha_{\mathrm{L}}}{\alpha_{\mathrm{S}}\cdot\alpha_{\mathrm{S}}}
\label{ng}
\end{equation}
where $\alpha_{\mathrm{L}}$ and $\alpha_{\mathrm{S}}$ denote, respectively, the long and the short roots
of the algebra $\mathfrak{g}$. In Appendix~\ref{secn:roots} we give our conventions on the root
systems, from which one can see that
\begin{equation}
\begin{aligned}
n_{\mathfrak{g}}&=2\quad\mbox{for}~\mathfrak{g}=B_r,C_r~\mbox{and}~F_4~,\\
n_{\mathfrak{g}}&=3\quad\mbox{for}~\mathfrak{g}=G_2~.
\end{aligned}
\end{equation}
Using this notation, we can write the prepotential for a theory with gauge algebra $\mathfrak{g}$ as
\begin{equation}
F^{(\mathfrak{g})}(\tau,a)= n_{\mathfrak{g}}\pi\ii\tau a^2+ f^{(\mathfrak{g})}(\tau,a)
\label{FG}
\end{equation}
where the first term represents the classical contribution. Here we have introduced the usual complex
combination of the Yang-Mills coupling $g$ and the $\theta$-angle
\begin{equation}
\tau=\frac{\theta}{2\pi}+\ii\,\frac{4\pi}{g^2}
\label{tau}
\end{equation}
and denoted by $a$ the vacuum expectation value of the scalar field $\varphi$ in the gauge vector multiplet
\begin{equation}
\langle\,\varphi \,\rangle = a = \mathrm{diag}(a_1,a_2,\cdots,a_r)~.
\label{a}
\end{equation}
Like in the ADE case \cite{Billo},
also the non-simply laced
quantum prepotential $f^{(\mathfrak{g})}$ can be conveniently expanded in the mass $m$ of the
adjoint hypermultiplet, namely
\begin{equation}
f^{(\mathfrak{g})}(\tau,a)=\sum_{n=1}^\infty  f_n^{(\mathfrak{g})}(\tau,a) ~,
\label{fgn}
\end{equation}
with $f_n^{(\mathfrak{g})}\sim m^{2n}$.  For $n\geq 2$ the coefficients $f_n^{(\mathfrak{g})}$ 
are functions of the coupling $\tau$ through the instanton counting parameter $q=\ee^{2\pi \ii \tau}$. 
On dimensional grounds, they are also homogeneous functions of degree $(2-2n)$ of the
vacuum expectation values $a$:
\begin{equation}
f_n^{(\mathfrak{g})}(\tau,\lambda a)= \lambda^{2-2n}\,f_n^{(\mathfrak{g})}(\tau,a)~.
\label{homoge}
\end{equation}
On the contrary, $f_1$ is independent of $\tau$ and is entirely given by the 1-loop contribution
\begin{equation}
f^{(\mathfrak{g})}_1(a,\Lambda)=f_1^{(\mathfrak{g}), {\mathrm{1-loop}}}(a,\Lambda) =\frac{m^2}{4} \sum_{\alpha\in\Psi}  \log\left(\frac{\alpha\cdot a}{\Lambda}\right)^2
\label{f1}
\end{equation}
where $\Lambda$ is the dynamically generated scale.

Since we are interested in the S-duality transformation of the prepotential (\ref{FG}),
we have to define how S-duality acts on the gauge coupling $\tau$ and on the vacuum expectation values $a$.
As discussed for example in \cite{Dorey:1996hx}\nocite{Argyres:2006qr}\,-\,\cite{Kapustin:2006pk}, 
in the non-simply laced theories one has
\begin{equation}
\tau\,\to\, \cS(\tau)=-\frac{1}{n_{\mathfrak{g}}\tau}~,
\label{stau}
\end{equation}
which replaces the usual $\tau\to-1/\tau$ transformation of the ADE models.
Furthermore, S-duality maps a theory with a gauge algebra $\mathfrak{g}$ to a
theory with a gauge algebra $\mathfrak{g}^{\!\vee}$ defined from $\mathfrak{g}$ by exchanging
(and suitably rescaling) the long and the short roots \cite{Goddard:1976qe}.
The correspondence between $\mathfrak{g}$ and $\mathfrak{g}^{\!\vee}$ is given in Tab.~1, where 
for $F_4$ and $G_2$, the $'$ in the second column means that the dual root systems are equivalent 
to the original ones up to a rotation.
\begin{table}[ht]
\begin{center}
\begin{tabular}{|c|c|}
\hline
~~$\mathfrak{g}$~~&~~$\mathfrak{g}^{\!\vee}$~~\\
\hline
\hline
$B_r$&$C_r$\\
$C_r$&$B_r$\\
$F_4$&$F'_4$\\
$G_2$&$G'_2$\\
\hline
\end{tabular}
\end{center}
\label{tab:Gdual}
\caption{The correspondence between a non-simply laced algebra $\mathfrak{g}$ and its GNO dual
$\mathfrak{g}^{\!\vee}$.}
\end{table}
Note that according to this definition, which is simply the $\mathcal{N}=2$ version of
the $\mathcal{N}=4$ S-duality rule 
\cite{Montonen:1977sn}\nocite{Goddard:1976qe,Girardello:1995gf,Dorey:1996hx,Argyres:2006qr}\,-\,\cite{Kapustin:2006pk}, the 
electric variables of the $\mathfrak{g}$ theory are dual to the magnetic variables of 
the $\mathfrak{g}^{\!\vee}$ theory. Moreover, using the roots given in Appendix~\ref{secn:roots}, from (\ref{f1}) 
it is easy to check that, up to
$a$-independent terms%
\footnote{Here and in the following we neglect all $a$-independent terms of the prepotential since they are irrelevant for the low-energy effective theory. These terms can always be absorbed by redefining 
the scale $\Lambda$ in one of the sides of (\ref{f1f1}).\label{footnote1}},
\begin{equation}
f_1^{(\mathfrak{g})}(a,\Lambda)=f_1^{(\mathfrak{g}^{\!\vee})}(a,\Lambda)~.
\label{f1f1}
\end{equation}

To define how S-duality acts on $a$, we have first to introduce the
dual variables $a_{\mathrm{D}}$. These are defined as the $a$-derivatives of the prepotential of the dual
$\mathfrak{g}^{\!\vee}$-theory, namely
\begin{equation}
\label{adnsl}
a_{\mathrm{D}} = \frac{1}{2n_{\mathfrak{g}}\pi\ii}\frac{\partial F^{(\mathfrak{g}^\vee)}}{\partial a}=
\tau 	\Big(a+\frac{\delta}{12n_{\mathfrak{g}}}\frac{\partial f^{(\mathfrak{g}^\vee)}}{\partial a}\Big)
\end{equation}
where, for later convenience, we introduce
\begin{equation}
\delta=\frac{6}{\pi\ii\tau}~.
\label{delta}
\end{equation}
The S-duality transformation (\ref{stau}) is represented by the $\mathrm{Sl}(2,\mathbb{R})$
element \cite{Dorey:1996hx,Kapustin:2006pk}
\begin{equation}
\cS=\begin{pmatrix}
0&-1/\sqrt{n_{\mathfrak{g}}} \\
\sqrt{n_{\mathfrak{g}}} &0
\end{pmatrix}
\label{Snsl}
\end{equation}
which, when acting on the periods, exchanges $a$ and $a_{\rm D}$; indeed
\begin{equation}
\begin{pmatrix}
a_{\mathrm{D}} \\
a
\end{pmatrix}\,\to\,
\begin{pmatrix}
0&-1/\sqrt{n_{\mathfrak{g}}} \\
\sqrt{n_{\mathfrak{g}}}&0
\end{pmatrix}\,\begin{pmatrix}
a_{\mathrm{D}} \\
a
\end{pmatrix}=\begin{pmatrix}
-a/\sqrt{n_{\mathfrak{g}}} \\
\sqrt{n_{\mathfrak{g}}}\, a_{\mathrm{D}}
\end{pmatrix}~.
\label{saad}
\end{equation}
Thus we have
\begin{equation}
\cS(a)=\sqrt{n_{\mathfrak{g}}} \,a_{\mathrm{D}}=\sqrt{n_{\mathfrak{g}}} \,\tau \Big(a+\frac{\delta}{12n_{\mathfrak{g}}}\frac{\partial f^{\mathfrak{g}^\vee}}{\partial a}\Big)~.
\label{sa}
\end{equation}
Using (\ref{stau}) and (\ref{sa}), the S-dual prepotential is therefore
\begin{equation}
\cS\big[F^{(\mathfrak{g})}\big] \,\equiv\,F^{(\mathfrak{g})} \left(\cS(\tau),\cS(a),\cS(\Lambda) \right)=
F^{(\mathfrak{g})} \left(\!-\ft{1}{n_{\mathfrak{g}}\tau},\sqrt{n_{\mathfrak{g}}} \,\tau \big(a+\ft{\delta}{12n_{\mathfrak{g}}}\ft{\partial f^{\mathfrak{g}^\vee}}{\partial a}\big),\cS(\Lambda) \right)~,
\label{dualF}
\end{equation}
where the modular transformation property of the scale $\Lambda$ will be determined shortly. 
In analogy with the ADE case \cite{Billo}, one can constrain the form of the prepotential
by requiring that $\cS\big[F^{(\mathfrak{g})}\big]$ be the Legendre transform of the prepotential
of the dual theory, namely
\begin{equation}
\cS\big[F^{(\mathfrak{g})}\big] =\mathcal{L}\big[F^{(\mathfrak{g}^{\!\vee})}\big]
\label{SL}
\end{equation}
where
\begin{equation}
\label{Legendre}
\begin{aligned}
\mathcal{L}\big[F^{(\mathfrak{g}^{\!\vee})}\big]
&\equiv F^{(\mathfrak{g}^{\!\vee})} - a\cdot\frac{\partial F^{(\mathfrak{g}^{\!\vee})}}{\partial a}
=-n_{\mathfrak{g}}\pi\ii\tau a^2 - a\cdot\frac{\partial f^{(\mathfrak{g}^{\!\vee})}}{\partial a}+f^{(\mathfrak{g}^{\!\vee})} ~.
\end{aligned}
\end{equation}
As it is clear from (\ref{SL}), S-duality is not a symmetry of the non-simply laced theories since
it relates a theory with gauge algebra $\mathfrak{g}$ to a theory with gauge algebra $\mathfrak{g}^{\!\vee}$ or viceversa. Nevertheless, it is powerful enough to constrain the structure of the prepotential.

In order to enforce (\ref{SL}), several conditions have to be satisfied. First of all, like in the ADE theories
\cite{Billo}, we will see that also here the prepotential coefficients
$f_n^{(\mathfrak{g})}$ (with $n\geq 2$) must depend on $\tau$ through quasi-modular forms. However, differently from the ADE models,
the modular group is now a subgroup of $\Gamma=\mathrm{Sl}(2,\mathbb{Z})$; more precisely
it is the congruence subgroup $\Gamma_0(n_{\mathfrak{g}})$ defined as%
\footnote{Note that the S-duality transformation (\ref{stau}) lies outside this subgroup.}
\begin{equation}
\Gamma_0(   n_{\mathfrak{g}} ) = \Bigg\{
\begin{pmatrix}
~a~&~b~ \\
~c~&~d~
\end{pmatrix} \in \Gamma~: ~c=0~~\mbox{mod}~n_{\mathfrak{g}}
\Bigg\}~.
\label{gamma02}
\end{equation}
If we denote by $\hat S=(^{0\,-1}_{1\,\,~0})$ and $\hat T=(^{1\,\,1}_{0\,\,1})$ 
the generators of $\Gamma$,
then $\Gamma_0(n_{\mathfrak{g}})$ is generated
by $\hat T$ and $\hat S\,{\hat T}^{n_{\mathfrak{g}}}\,\hat S$. The (quasi-)modular forms of
$\Gamma_0( n_{\mathfrak{g}})$ are known (see for instance \cite{Koblitz,Apostol}; see also \cite{sage} for a catalog and Appendix B of \cite{Angelantonj:2013eja} for a nice
compendium). They form a ring generated by the basic elements
\begin{equation}
\begin{aligned}
 \Big\{ E_2(\tau), H_2(\tau), E_4(\tau), E_6(\tau)\Big\}\quad\mbox{for}~n_{\mathfrak{g}}=2~, \\
 \Big\{ E_2(\tau), K_2(\tau), E_4(\tau), E_6(\tau) \Big\}\quad\mbox{for}~n_{\mathfrak{g}}=3~,
\label{modbasis}
\end{aligned}
\end{equation}
where $E_2$, $E_4$ and $E_6$ are the Eisenstein series of weight 2, 4 and 6 respectively, and $H_2$
and $K_2$ are modular forms of weight 2 defined by
\begin{equation}
\begin{aligned}
H_2(\tau) &=\frac{1}{2}\big(\theta_3^4(\tau)+\theta_4^4(\tau)\big)\\
&= 1+24 q + 24 q^2 + 96 q^3 + 24 q^4
+ 144 q^5+\cdots ~,\\
K_2(\tau) &=\left[\left(\frac{\eta^3(\tau)}{\eta(3\tau)}\right)^3+\left(\frac{3\eta^3(3\tau)}{\eta(\tau)}\right)^3\,\right]^{\frac{2}{3}}\\
&=1 + 12 q + 36 q^2 + 12 q^3 + 84 q^4 + 72 q^5
+\cdots~,
\label{h}
\end{aligned}
\end{equation}
where the $\theta$'s are the Jacobi $\theta$-functions and $\eta$ is the Dedekind $\eta$-function.
We refer to Appendix~\ref{secn:modular} for a summary of the main properties of these modular functions and
the Eisenstein series. Here we simply recall that all elements of the basis (\ref{modbasis}) are modular
forms of $\Gamma_0(n_{\mathfrak{g}})$, except for $E_2$ which is quasi-modular. Also under the  transformation (\ref{stau}), which does not belong to $\Gamma_0(n_{\mathfrak{g}})$,
all basis elements transform covariantly up to a factor of $\big(\sqrt{n_{\mathfrak{g}}}\tau\big)^w$ where $w$ is their modular weight, except for $E_2$ which behaves as
follows
\begin{equation}
E_2\big(\!-\ft{1}{n_\mathfrak{g}\tau}\big)= ~\begin{cases}
2\tau^2\Big[E_2(\tau)+\delta+H_2(\tau)\Big]\quad~\,\mbox{for}~n_{\mathfrak{g}}=2 ~,\phantom{\Bigg|} \\
3\tau^2\Big[E_2(\tau)+\delta+2K_2(\tau)\Big]\quad\mbox{for}~n_{\mathfrak{g}}=3~,
\end{cases}
\end{equation}
where $\delta$ is given in (\ref{delta}). This implies that under S-duality, up to the prefactors
of $\big(\sqrt{n_{\mathfrak{g}}}\tau\big)^2$ and the modular forms $H_2$ or $K_2$, any occurrence of $E_2$ is replaced by $E_2+\delta$. Since this ``anomalous" shift plays
a crucial r\^ole in the following, we will explicitly exhibit the $E_2$-dependence of the prepotential coefficients $f_n^{(\mathfrak{g})}$ by writing, for $n\geq 2$,
\begin{equation}
f_n^{(\mathfrak{g})}\big(\tau,a,E_2)~.
\end{equation}
On the other hand we leave implicit the dependence on the other modular forms to avoid clutter in the formulas.

Guided by the experience with the ADE theories \cite{Billo}, after some simple algebra
one can realize that in order to enforce (\ref{SL}) it is necessary that the
coefficients $f_n^{(\mathfrak{g})}$ behave as
\begin{equation}
f_n^{(\mathfrak{g})}\big(\!-\ft{1}{n_{\mathfrak{g}}\tau},a,
E_2(-\ft{1}{n_{\mathfrak{g}}\tau})\big)= \big(\sqrt{n_{\mathfrak{g}}}\,\tau\big)^{2n-2} \,
f_n^{(\mathfrak{g}^{\!\vee})}\big(\tau,a,E_2+\delta\big)
\label{fng}
\end{equation}
for $n\geq 2$. The prefactor $(\sqrt{n_{\mathfrak{g}}}\,\tau)^{2n-2}$ in the right hand side
is precisely the one that the (quasi-)modular forms of
$\Gamma_0(n_{\mathfrak{g}})$ of weight $(2n-2)$ acquire under the S-duality transformation (\ref{stau}).
Thus, we conclude that $f_n^{(\mathfrak{g})}$ must be a (quasi-)modular form of $\Gamma_0(n_{\mathfrak{g}})$ with weight $(2n-2)$.
Thanks to the homogeneity property (\ref{homoge}), it is possible to rewrite (\ref{fng}) as
\begin{equation}
f_n^{(\mathfrak{g})}\big(\!-\ft{1}{n_{\mathfrak{g}}\tau},\sqrt{n_\mathfrak{g}}\tau\,a,
E_2(-\ft{1}{n_{\mathfrak{g}}\tau})\big)=f_n^{(\mathfrak{g}^{\!\vee})}\big(\tau,a,E_2+\delta\big)~.
\label{fng1}
\end{equation}
For $n=1$, instead, we simply have to require that
\begin{equation}
f_1^{(\mathfrak{g})}(\sqrt{n_\mathfrak{g}}\tau\,a,\cS(\Lambda))=f_1^{(\mathfrak{g^{\!\vee}})}(a,\Lambda)~,
\label{f1g}
\end{equation}
which, in view of (\ref{f1f1}), implies
\begin{equation}
\cS(\Lambda)= \sqrt{n_\mathfrak{g}}\tau\,\Lambda~.
\label{Lambda1}
\end{equation}
Eq.s~(\ref{fng1}) and (\ref{f1g}) can be combined together into
\begin{equation}
f^{(\mathfrak{g})}\big(\!-\ft{1}{n_{\mathfrak{g}}\tau},\sqrt{n_\mathfrak{g}}\tau\,a,
E_2(-\ft{1}{n_{\mathfrak{g}}\tau}) ,  \sqrt{n_\mathfrak{g}}\tau      \Lambda \big)
=f^{(\mathfrak{g}^{\!\vee})}\big(\tau,a,E_2+\delta, \Lambda \big)~.
\label{dualf}
\end{equation}

We are now in the position of computing the S-dual prepotential. We have
\begin{equation}
\begin{aligned}
\mathcal{S}\big[F^{(\mathfrak{g})}\big]&=
F^{(\mathfrak{g})}\Big(\!-\ft{1}{n_{\mathfrak{g}}\tau},\sqrt{n_\mathfrak{g}}\tau\big(a+\ft{\delta}{12n_{\mathfrak{g}}}\ft{\partial f^{(\mathfrak{g}^{\!\vee})}}{\partial a}\big),
E_2(-\ft{1}{n_{\mathfrak{g}}\tau}) ,  \sqrt{n_\mathfrak{g}}\tau      \Lambda  \Big) \phantom{\Big|}\\
&=-n_{\mathfrak{g}}\pi\ii\tau a^2 - a\cdot\frac{\partial f^{(\mathfrak{g}^\vee)}}{\partial a}
-\frac{\delta}{24n_{\mathfrak{g}}}\,\frac{\partial f^{(\mathfrak{g}^\vee)}}{\partial a}\cdot
\frac{\partial f^{(\mathfrak{g}^\vee)}}{\partial a}\\
&~~~~+f^{(\mathfrak{g}^{\!\vee})}
\Big(\tau,a+\ft{\delta}{12n_{\mathfrak{g}}}\ft{\partial f^{(\mathfrak{g}^\vee)}}{\partial a},E_2+\delta,\Lambda\Big)~.
\end{aligned}
\label{SF}
\end{equation}
The second line is the S-dual of the classical prepotential,
while the third line gives the S-dual of the quantum prepotential $f^{(\mathfrak{g})}$ which has been
expressed in terms of  $f^{(\mathfrak{g}^{\!\vee})}$ using (\ref{dualf}).
Taylor expanding the right hand side (\ref{SF}) with respect to $\delta$, we get
\begin{eqnarray}
\mathcal{S}\big[F^{(\mathfrak{g})}\big]&=&-n_{\mathfrak{g}}\pi\ii\tau a^2 - a\cdot\frac{\partial f^{(\mathfrak{g}^\vee)}}{\partial a}+f^{(\mathfrak{g}^\vee)}
\big(\tau,a,E_2,\Lambda\big)\nonumber\\
&&+\delta\,\Big(\frac{\partial f^{(\mathfrak{g}^\vee)}}{\partial E_2}+\frac{1}{24\,n_{\mathfrak{g}}}
\frac{\partial f^{(\mathfrak{g}^\vee)}}{\partial a}\cdot
\frac{\partial f^{(\mathfrak{g}^\vee)}}{\partial a}\Big)\label{SFexp}\\
&&+\frac{\delta^2}{2}\,\Big(\frac{1}{144 n^2_{\mathfrak{g}}}
\frac{\partial f^{(\mathfrak{g}^\vee)}}{\partial a}\cdot
\frac{\partial^2 f^{(\mathfrak{g}^\vee)}}{\partial^2 a}\cdot
\frac{\partial f^{(\mathfrak{g}^\vee)}}{\partial a}+\frac{\partial^2 f^{(\mathfrak{g}^\vee)}}{\partial^2 E_2}+ \frac{1}{6n_{\mathfrak{g}}}\frac{\partial f^{(\mathfrak{g}^\vee)}}{\partial a}\cdot\frac{\partial^2 f^{(\mathfrak{g}^\vee)}}{\partial a\,\partial E_2}\Big)\nonumber\\
&&+\mathcal{O}(\delta^3)~.\nonumber
\end{eqnarray}
The first line in the right hand side reproduces the Legendre transform of the dual prepotential (\ref{Legendre}), so  in order to enforce the relation (\ref{SL}) the
$\delta$-dependent terms should vanish.
The cancellation of the term linear in $\delta$ implies
\begin{equation}
\frac{\partial f^{(\mathfrak{g}^{\!\vee})}}{\partial E_2}+\frac{1}{24\,n_{\mathfrak{g}}}
\frac{\partial f^{(\mathfrak{g}^{\!\vee})}}{\partial a}\cdot
\frac{\partial f^{(\mathfrak{g}^{\!\vee})}}{\partial a}=0~.
\label{diffeq}
\end{equation}
We have written this modular anomaly equation for the $\mathfrak{g}^{\!\vee}$-theory, but it clearly holds also for the dual $\mathfrak{g}$-theory.
The cancellation of the $\delta^2$-term follows from differentiating (\ref{diffeq}) with respect to $E_2$.
By taking further $E_2$ derivatives of this differential equation one
can check that also the higher order terms in the $\delta$-expansion vanish.

Summarizing, the S-duality symmetry relation (\ref{SL}) requires that the mass expansion coefficients $f_n^{(\mathfrak{g})}$ of the quantum prepotential are quasi-modular forms of the congruence
subgroup $\Gamma_0(n_{\mathfrak{g}})\subset\mathrm{Sl}(2,\mathbb{Z})$ with weight $(2n-2)$
satisfying the recursion relations
\begin{equation}
\frac{\partial f_n^{(\mathfrak{g})}}{\partial E_2}=-\frac{1}{24\,n_{\mathfrak{g}}}\sum_{\ell=1}^{n-1}\, \frac{\partial f_\ell^{(\mathfrak{g})}}{\partial a}\cdot
\frac{\partial f_{n-\ell}^{(\mathfrak{g})}}{\partial a}~.
\label{recursionG}
\end{equation}
Moreover, since the S-duality map (\ref{fng1}) exchanges the algebras $\mathfrak{g}$ and
$\mathfrak{g}^{\!\vee}$, the prepotential coefficients $f_n^{(\mathfrak{g})}$
should depend on the vacuum expectation values $a$ through the long and short roots
of $\mathfrak{g}$ in a way that is compatible with (\ref{dualf}). This is a highly non-trivial requirement. However, as we will see later on, the explicit evaluation of the
first few instanton corrections in the $B_r$ and $C_r$ theories reveals
that the $a$-dependent part to the prepotential can be written in terms of the following
two basic sums
\begin{equation}
\begin{aligned}
L_{n;m_1\,m_2\cdots\, m_\ell}&=\sum_{\alpha\in\Psi_{\mathrm{L}}}~\sum_{\beta_1\not=\beta_2\not=\cdots\beta_\ell\in\Psi(\alpha)}
\frac{1}{(\alpha\cdot a)^n(\beta_1\cdot a)^{m_1}(\beta_2\cdot a)^{m_2}
\cdots(\beta_\ell\cdot a)^{m_\ell}}~,\\
S_{n;m_1\, m_2\cdots\,m_\ell}&=\sum_{\alpha\in\Psi_{\mathrm{S}}}~\sum_{\beta_1\not=\beta_2\not=\cdots\beta_\ell\in\Psi^\vee(\alpha)}
\frac{1}{(\alpha\cdot a)^n(\beta^\vee_1\cdot a)^{m_1}(\beta^\vee_2\cdot a)^{m_2}
\cdots(\beta^\vee_\ell\cdot a)^{m_\ell}}~,
\end{aligned}
\label{sumsLS}
\end{equation}
where we have denoted by $\Psi_{\mathrm{L}}$ and $\Psi_{\mathrm{S}}$, respectively, the sets
of long and short roots of $\mathfrak{g}$, and defined for any root $\alpha$
\begin{equation}
\begin{aligned}
\Psi(\alpha)&=\left\{\beta\in\Psi \, : \,\alpha^{\!\vee}\cdot\beta=1\right\}~,\\
\Psi^{\vee}(\alpha)&=\left\{\beta\in\Psi \, :\,\alpha\cdot\beta^{\vee}=1\right\}~,
\end{aligned}
\end{equation}
with $\alpha^{\!\vee}$ denoting the coroot of $\alpha$ (see Appendix \ref{secn:roots} for details).
The sums (\ref{sumsLS}) are a generalization for the non-simply laced groups of the sums $C_{n;m_1\cdots}$
introduced in \cite{Billo} for the ADE series. With an abuse of language, we will often refer to
$L_{n;m_1\cdots}$ and $S_{n;m_1\cdots}$ as the ``long" and ``short" sums, respectively, since the first
factors in the denominators involve long and short roots. Using the properties of the root systems
(see Appendix~\ref{secn:roots}) it is not difficult to show that
\begin{equation}
\begin{aligned}
L_{n;m_1\cdots m_\ell}^{(\mathfrak{g})}&= \Big(\frac{1}{\sqrt{n_{\mathfrak{g}}}}\Big)^{n+m_1+\cdots+m_\ell}
\,S_{n;m_1\cdots m_\ell}^{(\mathfrak{g}^{\!\vee})}~,\\
S_{n;m_1\cdots m_\ell}^{(\mathfrak{g})}&= \big(\sqrt{n_\mathfrak{g}}\big)^{n+m_1+\cdots+m_\ell}
\,L_{n;m_1\cdots m_\ell}^{(\mathfrak{g}^{\!\vee})}~.
\end{aligned}
\label{LSa}
\end{equation}
These are precisely the desired maps.

Showing that all this construction can be explicitly realized and proved will be the subject of the remainder
of this paper.

\section{Multi-instanton calculations for the $B_r$ and $C_r$ theories}
\label{secn:multiinst}

In this section we discuss multi-instanton calculations in $\mathcal{N}=2^\star$ theories with the classical
gauge algebras $B_r$ and $C_r$, using the methods of equivariant localization \cite{Nekrasov:2002qd}\nocite{Nekrasov:2003rj,Bruzzo:2002xf}\,-\,\cite{Fucito:2004ry}. This is necessary
to obtain explicit expressions for the prepotential coefficients $f_n$, order by order
in the instanton expansion, and verify that indeed they can be resummed into quasi-modular forms
of $\Gamma_0(2)$ as anticipated in the previous section. Even if some multi-instanton calculations for orthogonal and symplectic theories with adjoint matter have
already been considered in the literature (see for example \cite{Shadchin:2004yx,Marino:2004cn}), we present a brief
discussion here in order to be as self-contained as possible, and also to fix some details and subtleties
that have been overlooked, but which are important for the explicit calculations.

We recall that the instanton moduli space for theories with orthogonal and symplectic gauge groups
can be engineered using systems of $N$ D3 branes and $k$ D(-1) branes living on top on an orientifold
O3 plane in Type IIB string theory \cite{Fucito:2004ry}. In this set-up the instanton moduli are realised in terms of the lowest modes of open strings with at least an end-point on a D(-1) brane
\cite{Witten:1995gx}\nocite{Douglas:1995bn,Green:2000ke,Billo:2002hm}\,-\,\cite{Billo:2006jm}, whose spectrum can be obtained
from that of the parent $\mathrm{U}(N)\times \mathrm{U}(k)$ theory after an orientifold projection.

\subsection{Multi-instantons for the $B_r=\mathrm{so}(2r+1)$ theories}
 \label{secn:multiBr}

The moduli space of the SO$(2r+1)$ gauge theory can be found from that of the U$(2r+1)$ theory
by quotienting it with $\Omega\, I$, where $\Omega$ is the parity operator that changes
the open string orientation, and $I$ is the operator that reflects those fields transforming as anti-chiral spinors, namely in the fundamental representation of the left-moving $SU(2)_{\mathrm{L}}$ subgroup
of the Lorentz group.  As a result of this projection,
keeping only the anti-symmetric combinations,
the symmetry group of the D3/D(-1) brane system
reduces to $\mathrm{SO}(2r+1)\times \mathrm{Sp}(2k)$, see for example \cite{Fucito:2004ry}. The invariant components
under $\Omega\,  I$  are listed in Tab.~2.
 \begin{table}[htb]
\begin{center}
\begin{tabular}{|c|c|c|c|}
\hline
\begin{small} $ (\phi,\psi) $ \end{small}
&\begin{small} $(-1)^{F_\phi}$ \end{small}
&\begin{small} $\mathrm{SO}(2r+1)\times\mathrm{Sp}(2k)$
\end{small}
&\begin{small}  $ \lambda_\phi\phantom{\Big|}$ \end{small}
\\
\hline\hline
$(B_{\alpha\dot\alpha},M_{\alpha \dot a} )$ & $+\phantom{\Big|}$ & $\bigl(\bone,\Yasymm\bigr)$
&   $ \chi_{ij}+\epsilon_1,\,\chi_{ij}+\epsilon_2$\\
$(B_{ a \dot a},M_{\dot\alpha  a} )$ & $+\phantom{\Big|}$& $\bigl(\bone,\Ysymm\bigr)$
&   $ \chi_{ij}+\epsilon_3,\,\chi_{ij}+\epsilon_4$\\
$(N_{(\dot\alpha\dot b)},D_{(\dot\alpha\dot \beta)})$ & $-\phantom{\Big|}$
& $\bigl(\bone,\Ysymm\bigr)  $ & $ \sqrt{\chi_{ij}},\, \chi_{ij}+\epsilon_1+\epsilon_2  $\\
$( \bar \chi, N)$ & $+\phantom{\Big|}$
& $\bigl(\bone,\Ysymm\bigr)  $ & $ \sqrt{\chi_{ij}}   $\\
$(N_{\alpha a},D_{\alpha a})$ & $-\phantom{\Big|}$
& $\bigl(\bone,\Yasymm\bigr)  $ & $ \chi_{ij}+\epsilon_1+\epsilon_3,\, \chi_{ij}+\epsilon_1+\epsilon_4  $\\
$(w_{\dot \alpha},\mu_{\dot a})$ & $+\phantom{\Big|}$
& $\bigl(\Yfund,  {\Yfund}\bigr)  $
   & $ \chi_i-\varphi_u+\ft{\epsilon_1+\epsilon_2}{2} $  \\
   $(h_{a},\mu_{a})$ & $-\phantom{\Big|}$
& $\bigl( \Yfund,  {\Yfund} \bigr)  $
   &   $ \chi_i-\varphi_u+\ft{\epsilon_3-\epsilon_4 }{2}  $  \\
\hline
\end{tabular}
\caption{Instanton moduli for the SO$(2r+1)$ gauge theory. The columns display, respectively,
the moduli in a ADHM-like notation organized as supersymmetric pairs, their statistics,
their transformation properties with respect to the gauge and instanton symmetry groups and finally
$Q^2$-eigenvalues $\lambda_\phi$, where $Q$ is the supersymmetry charge used in the localization approach.  See also \cite{Billo:2012st} where similar tables have been given for other brane systems.}
\end{center}
\label{tabson}
\end{table}

The most relevant information is contained in the last column of the above table:
there the $\chi_i$'s (with $i=1,\cdots,k$) are the unpaired bosonic moduli representing the D(-1) positions,
  $\varphi_u$'s (with $u=1,\cdots,r$) are the D3 positions and are related to the
vacuum expectation values $a_u$; the notation $\chi_{ij}$ stands for $\chi_i-\chi_j$, and finally
$\epsilon_{1,2,3,4}$ are the deformation parameters of the $SO(4)\times SO(4)$ symmetry. Collecting all eigenvalues $\lambda_\phi$, one finds that the $k$-instanton partition function is given by
\begin{equation}
Z_k=
\oint\prod_{i=1}^k\frac{d\chi_i}{2\pi\ii}~z_k^{\mathrm{gauge}}\,z_k^{\mathrm{matter}}
\label{Zkson}
\end{equation}
where
\begin{align}
z_k^{\mathrm{gauge}}=&\, \,
\frac{(-1)^k}{2^k\,k!} \left( \frac{\epsilon_1+\epsilon_2}{\epsilon_1\epsilon_2}\right)^k
\frac{\Delta(0)\,\Delta(\epsilon_1+\epsilon_2)}{\Delta(\epsilon_1)\,\Delta(\epsilon_2)}\,\prod_{i=1}^k
\frac{ 4\chi_i^2 \,\big( 4\chi_i^2-(\epsilon_1+\epsilon_2)^2\big)}{P\big(\chi_i+\frac{\epsilon_1+\epsilon_2}{2}\big)P\big(\chi_i-\frac{\epsilon_1+\epsilon_2}{2}\big)}
\notag~,\\
\notag\\
z_k^{\mathrm{matter}}=&\,\left( \frac{ (\epsilon_1+\epsilon_3) (\epsilon_1+\epsilon_4)}{\epsilon_3 \epsilon_4}\right)^k
\frac{\Delta\big(\epsilon_1+\epsilon_3 \big)\Delta\big(\epsilon_1+\epsilon_4 \big)}  {\Delta\big(\epsilon_3 \big)\Delta\big(\epsilon_4 \big)}
\prod_{i=1}^k
\frac{P\big(\chi_i+\frac{\epsilon_3-\epsilon_4}{2} \big)P\big(\chi_i- \frac{\epsilon_3-\epsilon_4}{2} \big)}{\big( 4 \chi_i^2-\epsilon_3^2 \big)\big( 4 \chi_i^2-\epsilon_4^2 \big)}~,\notag
\end{align}
with
\begin{equation}
\begin{aligned}
P(x) &= x \prod_{u=1}^r\big(x^2-\varphi_u^2)~,  \\
\Delta(x)&=\prod_{i<j}^k\big((\chi_i-\chi_j)^2-x^2)\big)\big((\chi_i+\chi_j)^2-x^2\big)~.
\end{aligned}
\end{equation}
The integrals in (\ref{Zkson}) are computed by closing the contours in the upper-half complex $\chi_i$-planes after giving the deformation parameters an imaginary part with the following prescription
\begin{equation}
\mathrm{Im}(\epsilon_4)\gg \mathrm{Im}(\epsilon_3)\gg\mathrm{Im}(\epsilon_2)\gg\mathrm{Im}(\epsilon_1)> 0~.
\label{prescription}
\end{equation}
This choice allows us to unambiguously compute all integrals in (\ref{Zkson}) and to obtain the
instanton partition function
\begin{equation}
Z_{\mathrm{inst}}=1+\sum_{k=1}q^k \,Z_k~,
\end{equation}
where $q=\ee^{2\pi\ii\tau}$. At the end of the computation, we have to set
\begin{equation}
\varphi_u=\sqrt{2}\,a_u~, \qquad
\epsilon_3=m-\frac{\epsilon_1+\epsilon_2}{2}~,   \qquad
\epsilon_4=-m-\frac{\epsilon_1+\epsilon_2}{2}
\label{mass34}
\end{equation}
in order to express the result in terms of the vacuum expectation values $a_u$ and
the adjoint hypermultiplet mass $m$ in the normalization used in the previous section.
Finally, the instanton prepotential of the $\mathcal{N}=2^\star$ theory is given by
\begin{equation}
F_{\mathrm{inst}}= \lim_{\epsilon_1,\epsilon_2\to 0}\Big(-\epsilon_1\epsilon_2\,\log Z\Big)=
\sum_{k=1} q^k\,F_k~.
\end{equation}

\subsubsection*{1-instanton}
At $k=1$ there is just one integral to compute and one can easily see that the poles of the
integrand in (\ref{Zkson}) are located at
\begin{equation}
\chi_1= \left\{ \pm \,\varphi_u+\frac{\epsilon_1+\epsilon_2}{2}~,~\frac{\epsilon_3}{2}  ~,~\frac{\epsilon_4}{2} \right\}   \quad\mbox{for}~u=1,\cdots, r~.
 \end{equation}
Therefore the 1-instanton prepotential $F_{k=1}$ can be written as
\begin{equation}
\begin{aligned}
F_{k=1}&=\lim_{\epsilon_1,\epsilon_2\to 0}\big( -\epsilon_1 \epsilon_2 \, Z_{1}\big)\\
&=\lim_{\epsilon_1,\epsilon_2\to0}\left(\sum_{u=1}^r f_{+\varphi_u+\ft{\epsilon_1+\epsilon_2}{2}}+
\sum_{u=1}^r f_{-\varphi_u+\ft{\epsilon_1+\epsilon_2}{2}}\,+\,f_{\ft{\epsilon_3}{2}}\,+\,f_{\ft{\epsilon_4}{2}}
\right)
\end{aligned}
\label{Fk=1}
\end{equation}
where
\begin{equation}
\begin{aligned}
f_{ \pm \varphi_u+\ft{\epsilon_1+\epsilon_2}{2}} &=- (\epsilon_1+\epsilon_3) (\epsilon_1+\epsilon_4) \,
      \frac{(\pm 2\varphi_u+\epsilon_1+\epsilon_2)(\pm \varphi_u-\epsilon_3 )(\pm \varphi_u-\epsilon_4)}{(\pm \varphi_u+\epsilon_1+\epsilon_2-\epsilon_3)( \pm 2\varphi_u+\epsilon_1+\epsilon_2-\epsilon_4) \varphi_u }\\
&~~~~~~~~~\quad\times    \prod_{v\neq u}  \frac{ \big( (\pm \varphi_u-\epsilon_3)^2-\phi_v^2\big)
\big( (\pm \varphi_u-\epsilon_4)^2-\varphi_v^2\big)}{(\varphi_u^2-\varphi_v^2) \big(
(\pm \varphi_u+\epsilon_1+\epsilon_2)^2-\varphi_v^2\big) }~,\\
  f_{ \ft{\epsilon_3}{2} } &= -\frac{(\epsilon_1+\epsilon_3) (\epsilon_1+\epsilon_4) (2\epsilon_3-\epsilon_4)}{(\epsilon_3-\epsilon_4)}
      \prod_{u=1}^r \frac{\varphi_u^2-(2\epsilon_3-\epsilon_4)^2}{\varphi_u^2 (\epsilon_3-\epsilon_1-\epsilon_2)^2} ~,\\
    f_{ \ft{\epsilon_4}{2} } &= -\frac{(\epsilon_1+\epsilon_3) (\epsilon_1+\epsilon_4)  (2\epsilon_4-\epsilon_3)}{8  (\epsilon_3-\epsilon_4) }
      \prod_{u=1}^r  \frac{ (2\epsilon_4-\epsilon_3)^2-a_u^2}{(\epsilon_4-\epsilon_1-\epsilon_2)^2-\varphi_u^2}  ~.
\end{aligned}
\end{equation}
Inserting these expressions in (\ref{Fk=1}) and using (\ref{mass34}), for the first few
algebras of the $B_r$ series we obtain
\begin{subequations}
\begin{align}
F_{k=1}^{(B_1)}&= -\frac{5m^2}{8}~,\\
F_{k=1}^{(B_2)}&= -\frac{13m^2}{8}+\frac{2m^4\big(a_1^2+a_2^2\big)}{\big(a_1^2-a_2^2\big)^2}
-\frac{2m^6}{\big(a_1^2-a_2^2\big)^2}~,\\
F_{k=1}^{(B_3)}&=-\frac{21m^2}{8}+\frac{2 m^4 \left(a_1^4-a_2^2 a_1^2-a_3^2 a_1^2+a_2^4+a_3^4-a_2^2 a_3^2\right)}{\left(a_1^2-a_2^2\right)^2 \left(a_1^2-a_3^2\right)^2 \left(a_2^2-a_3^2\right)^2}
\notag\\
&\qquad\qquad\times\left(a_2^2 a_1^4+a_3^2 a_1^4+a_2^4 a_1^2+a_3^4 a_1^2-6
   a_2^2 a_3^2 a_1^2+a_2^2 a_3^4+a_2^4 a_3^2\right)+\cdots
\end{align}
\label{fk1so}
\end{subequations}
where the ellipses stand for terms with higher powers of the mass whose explicit expressions can be
obtained from (\ref{Fk=1}) in a straightforward way but rapidly become quite cumbersome.
We have checked (up to $B_5$) that our 1-instanton results exactly match those derived long ago in \cite{Ennes:1999fb} using very different methods.

\subsubsection*{Up to two instantons}
At $k=2$ one has to compute two integrals to obtain the instanton partition function and hence the
prepotential $F_{k=2}$. The procedure we have outlined above is straightforward to implement, and with the prescription
(\ref{prescription}) no ambiguity arises. For $B_1$ and $B_2$ we obtain the following 2-instanton contributions
\begin{subequations}
\begin{align}
F_{k=2}^{(B_1)}&= -\frac{23m^2}{16}+\frac{m^4}{a_1^2}~,\\
F_{k=2}^{(B_2)}&= -\frac{47m^2}{16}+\frac{m^4\big(a_1^6+5a_1^4 a_2^2+5a_1^2a_2^4+a_2^6\big)}{a_1^2a_2^2 \big(a_1^2-a_2^2\big)^2}\notag\\
&\qquad-\frac{2m^6 \big(a_1^8+5a_1^6a_2^2+12a_1^4a_2^4
+5a_1^2a_2^6+a_2^8\big)}{a_1^2 a_2^2 \big(a_1^2-a_2^2\big)^4}+\cdots
\end{align}
\label{Fk=2}
\end{subequations}
where again the ellipses stand for higher order mass terms.
We refrain from writing the explicit expressions of $F_{k=2}$ for other orthogonal
algebras since they are quite involved.
However, if we use the ``long" and ``short" sums defined in (\ref{sumsLS}), it is possible to write all the $k=1,2$ results in a very compact and simple way. Indeed, we have%
\footnote{We neglect again all $a$-independent terms, see footnote \ref{footnote1}.}
\begin{equation}
\begin{aligned}
F_{k=1}^{(B_r)}&= m^4 L_2+\frac{m^6}{2}L_{2;11}+ \frac{m^8}{24}L_{2;1111}+\cdots~,\\
F_{k=2}^{(B_r)}&= m^4 \big(3L_2+S_2\big)-m^6\Big(6L_4 - 3L_{2;11}+\frac12 S_{2;11}\Big)\\
&\qquad+
m^8\Big(\frac{5}{2}L_6+6L_{4;2}+\frac{1}{2}L_{3;3}+\frac{1}{2}L_{2;1111}+\frac{1}{16}S_{3;3}+
\frac{1}{24}S_{2;1111}\Big)+\cdots~.
\end{aligned}
\end{equation}
These formulas, which we have explicitly verified up to $B_5$, clearly show the advantage of organize
the multi-instanton results in terms of  the ``long" and ``short" sums that fully exploit the
algebraic properties of the root system of the gauge algebra.

\subsection{Multi-instantons for the $C_r=\mathrm{sp}(2r)$ theories}
\label{secn:multiCr}

The above analysis can be easily extended to the symplectic series $C_r$. The instanton moduli space of
the $\mathcal{N}=2^\star$ Sp$(2r)$ gauge theory is found from that of U$(2r)$ after quotienting
by $\Omega\, I$. As a result of this projection, keeping only the symmetric combinations,
the symmetry of the D3/D(-1) brane system reduces
to $\mathrm{Sp}(2r)\times SO(k)$ where $k=2K+\nu$ with $\nu=0$ or $1$ for $k$ even or odd respectively
\cite{Fucito:2004ry}.
The invariant components under $\Omega\,I$ are listed in Tab.~3.
\begin{table}[htb]
\begin{center}
\begin{tabular}{|c|c|c|c|}
\hline
\begin{small} $ (\phi,\psi) $ \end{small}
&\begin{small} $(-)^{F_\phi}$ \end{small}
&\begin{small} $\mathrm{Sp}(2r) \times \mathrm{SO}(k)$
\end{small}
&\begin{small}  $ \lambda_\phi\phantom{\Big|} $ \end{small}
\\
\hline\hline
$(B_{\alpha\dot\alpha},M_{\alpha \dot a} )$ & $+\phantom{\Big|}$ & $\bigl(\bone,\Ysymm\bigr)$
&   $ \chi_{ij}+\epsilon_1,\,\chi_{ij}+\epsilon_2$\\
$(B_{ a \dot a},M_{\dot\alpha  a} )$ & $+\phantom{\Big|}$ & $\bigl(\bone,\Yasymm\bigr)$
&   $ \chi_{ij}+\epsilon_3,\,\chi_{ij}+\epsilon_4$\\
$(N_{\dot\alpha\dot b},D_{\dot\alpha\dot \beta})$ & $-\phantom{\Big|}$
& $\bigl(\bone,\Yasymm\bigr)  $ & $ \chi_{ij},\, \chi_{ij}+\epsilon_1+\epsilon_2  $\\
$(N_{\alpha a},D_{\alpha a})$ & $-\phantom{\Big|}$
& $\bigl(\bone,\Ysymm\bigr)  $ & $ \chi_{ij}+\epsilon_1+\epsilon_3, \,\chi_{ij}+\epsilon_1+\epsilon_4  $\\
$(w_{\dot \alpha},\mu_{\dot a})$ & $+\phantom{\Big|}$
& $\bigl(\Yfund, \Yfund\bigr)  $
   & $ \chi_i-\varphi_u+\ft{\epsilon_1+\epsilon_2}{2}    $  \\
   $(h_{a},\mu_{a})$ & $-\phantom{\Big|}$
& $\bigl( \Yfund, \Yfund\bigr)  $
   &   $ \chi_i-\varphi_u+\ft{\epsilon_3-\epsilon_4 }{2}  $  \\
\hline
\end{tabular}
\caption{Instanton moduli for the $\mathcal{N}=2^\star$ Sp$(2r)$ gauge theory. As before,
the columns display, respectively, the moduli in a ADHM-like notation organized as supersymmetric pairs,
their statistics, their transformation properties with respect to the gauge and instanton symmetry
groups and finally $Q^2$-eigenvalues $\lambda_\phi$, where $Q$ is the supersymmetry charge
used in the localization approach.}
\end{center}
\label{tabspn}
\end{table}
Collecting the eigenvalues $\lambda_\phi$ from the last column of the above table, one finds that the
instanton partition function is now given by
\begin{equation}
Z_k=
\oint\prod_{i=1}^K\frac{d\chi_i}{2\pi\ii}~z_k^{\mathrm{gauge}}\,z_k^{\mathrm{matter}}
\label{Zkspn}
\end{equation}
where
\begin{align}
&z_k^{\mathrm{gauge}}=\frac{(-1)^k}{2^{k+\nu} \,k!}\,
\frac{ (\epsilon_1+\epsilon_2)^k}{ (\epsilon_1\epsilon_2)^{k+\nu}}
\,\frac{\Delta(0)\,\Delta(\epsilon_1+\epsilon_2 )}{\Delta(\epsilon_1)\,\Delta(\epsilon_2)}\,
   \frac{1}{P\left(\ft{\epsilon_1+\epsilon_2}{2} \right)^\nu}      \notag\\
& \quad\qquad\quad \times~\prod_{i=1}^K
\frac{1}{P\big(\chi_i+\frac{\epsilon_1+\epsilon_2 }{2}\big)P\big(\chi_i-\frac{\epsilon_1+\epsilon_2 }{2}\big)  (4\chi_i^2-\epsilon_1^2)   \big( 4\chi_i^2-\epsilon_2^2\big)}\notag\\
\notag\\
&z_k^{\mathrm{matter}}=\, \frac{ \big((\epsilon_1+\epsilon_3) (\epsilon_1+\epsilon_4)\big)^{k+\nu}}{  \left( \epsilon_3 \epsilon_4 \right)^k }
\frac{\Delta\big(\epsilon_1+\epsilon_3 \big)\Delta\big(\epsilon_1+\epsilon_4 \big)  }  {\Delta\big(\epsilon_3 \big)\Delta\big(\epsilon_4 \big)  }
 P\left(\ft{\epsilon_3-\epsilon_4}{2}\right)^\nu  \notag\\
&\quad\qquad\quad \times~\prod_{i=1}^K
P\left(\chi_i+\ft{\epsilon_3-\epsilon_4}{2} \right)P\left(\chi_i- \ft{\epsilon_3-\epsilon_4}{2} \right)
\big( 4 \chi_i^2-(\epsilon_1+\epsilon_3)^2 \big)\big( 4 \chi_i^2-(\epsilon_1+\epsilon_4)^2 \big) ~.
\notag
\end{align}
with
\begin{equation}
 \begin{aligned}
P(x) &=  \prod_{u=1}^r\big(x^2-\varphi_u^2)~,  \\
\Delta(x)&=\prod_{i<j}^K\big(x^2-(\chi_i-\chi_j)^2)\big)\big( x^2-(\chi_i+\chi_j)^2\big)
 \prod_{i=1}^K \big( x^2-\chi_i^2\big)^\nu   ~.
\end{aligned}
 \end{equation}
The integrals in (\ref{Zkspn}) are computed again by closing the contours in the upper-half complex $\chi_i$-planes with the prescription (\ref{prescription}). At the end of the computation we should make the
 substitutions
 \begin{equation}
\varphi_u=a_u~, \qquad
\epsilon_3=m-\frac{\epsilon_1+\epsilon_2}{2}~,   \qquad
\epsilon_4=-m-\frac{\epsilon_1+\epsilon_2}{2}
\label{mass34a}
\end{equation}
in order to write the result in terms of the physical parameters of the gauge theory in the normalizations of
Section~\ref{secn:sduality}. {From} the partition function we can derive the instanton contributions of the
prepotential along the same lines discussed for the orthogonal groups.

\subsubsection*{1-instanton}
 For $k=1$, i.e. $K=0$ and $\nu=1$, there is no integral to be done. The prepotential following from (\ref{Zkspn}) in this case is simply
\begin{equation}
\begin{aligned}
F_{k=1}&=\lim_{\epsilon_1,\epsilon_2\to 0}\big( -\epsilon_1 \epsilon_2 \, Z_1\big)
\\
&=\lim_{\epsilon_1,\epsilon_2\to0}\Big[\frac12(\epsilon_1+\epsilon_3) (\epsilon_1+\epsilon_4) \,
    \prod_{u=1}^r\frac{4\, \varphi_u^2-(\epsilon_3-\epsilon_4 )^2}{4\, \varphi_u^2-(\epsilon_1+\epsilon_2)^2}\Big]~.
\end{aligned}
 \end{equation}
In particular for the first few symplectic algebras, using (\ref{mass34a}) we find
\begin{subequations}
\begin{align}
F_{k=1}^{(C_1)}&= -\frac{m^2}{2}+ \frac{m^4}{2a_1^2}~,\\
F_{k=1}^{(C_2)}&= -\frac{m^2}{2}+\frac{m^4\big(a_1^2+a_2^2\big)}{2a_1^2a_2^2}
-\frac{m^6}{2a_1^2a_2^2}~,\\
F_{k=1}^{(C_3)}&=-\frac{m^2}{2}+
\frac{m^4\big(a_1^2 a_2^2+a_3^2 a_2^2+a_1^2 a_3^2\big)}{2a_1^2 a_2^2a_3^2}
-\frac{m^6\big(a_1^2+a_2^2+a_3^2}{2 a_1^2 a_2^2 a_3^2}+\frac{m^8}{2 a_1^2 a_2^2 a_3^2}~.
\end{align}
\label{fk1sp}
\end{subequations}
We have checked (up to $C_5$) that our 1-instanton results exactly match those derived
in \cite{Ennes:1999fb} using very different methods.

\subsubsection*{Up to four instantons}
For the symplectic algebras one can push the calculation of the partition function and the prepotential to higher instanton numbers with relatively little effort. Indeed, for $k=2$ and $k=3$ one has to compute just one
integral, while there are only two integrals to compute for $k=4$ and $k=5$. As an example, we now
write the first multi-instanton terms of the prepotential for the Sp(4) theory that we have obtained with these methods:
\begin{subequations}
\begin{align}
F_{k=2}^{(C_2)}&=-\frac{19m^2}{8}+\frac{m^4\big(3a_1^6+5a_1^4a_2^2+5a_1^2a_2^4+3a_2^6\big)}{2a_1^2a_2^2\big(a_1^2-a_2^2\big)^2}\\
&~~
-\frac{m^6\big(3a_1^8+6a_1^6a_2^2 +14a_1^4a_2^4 +6a_1^2 a_2^6 +3a_2^8\big)}
{4a_1^4 a_2^4 \big(a_1^2-a_2^2\big)^2}+\cdots\notag\\
F_{k=3}^{(C_2)}&= -\frac{2m^2}{3} +\frac{2m^4 \big(a_1^2+a_2^2\big)}{a_1^2a_2^2}
-\frac{2m^6 \big(2a_1^4+3a_1^2a_2^2+2a_2^4\big)}{a_1^4 a_2^4}+\cdots
\\
F_{k=4}^{(C_2)}&=-\frac{53m^2}{16}  +\frac{m^4 \big( 7a_1^6+17a_1^4 a_2^2+17a_1^2 a_2^4
+7 a_2^6\big)}{2 a_1^2 a_2^2 \big(a_1^2-a_2^2\big)^2} \\
&~~-\frac{m^6\big(45a_1^{12}-124 a_1^{10}a_2^2 +379 a_1^8a_2^4 +168 a_1^6a_2^6
   +379 a_1^4a_2^8 -124 a_1^2a_2^{10} +45 a_2^{12}}{4 a_1^4 a_2^4\big(a_1-a_2\big)^4
   \big(a_1+a_2\big)^4}
+\cdots\notag
\end{align}
\end{subequations}
where the ellipses stand for terms with higher powers of $m$.
Clearly the explicit formulas become more and more involved for symplectic groups of higher rank, and quickly
cease to be useful. However, if we use the ``long" and ``short" sums (\ref{sumsLS}),
we can write quite compact
expressions which are valid for all $C_r$'s. Indeed, neglecting again the $a$-independent terms, we find
\begin{subequations}
\begin{align}
F_{k=1}^{(C_r)}&=m^4\,L_2+\frac{m^6}{2}\,L_{2;11}+\frac{m^8}{24}\,L_{2;1111}+\ldots~,\label{Fk1}\\
F_{k=2}^{(C_r)}&=m^4\big(3L_2+S_2\big)-m^6\left(6L_4-3L_{2;11}-\frac{1}{2}S_{2;11}\right)\notag\\
&~~~~~+
m^8\left(\frac{5}{2} L_6+6 L_{4;2}+ L_{3;3}
+\frac{1}{2}L_{2;1111}+\frac{1}{24}S_{2;1111}\right)+\cdots~,\\
F_{k=3}^{(C_r)}&=4m^4\,L_2-m^6\big(32L_4-6L_{2;11}\big)\notag\\
&~~~~~+m^8\left(80L_6+48L_{4;2}+8L_{3;3}+\frac{3}{2}L_{2;1111}\right)+\cdots~,\\
F_{k=4}^{(C_r)}&=m^4\big(7L_2+3S_2\big)-m^6\big(90L_4-14L_{2;11}+6S_4-3S_{2;11}\big)\notag\\
&~~~~~+
m^8\left(\frac{1095}{2} L_6+180 L_{4;2}+ 34L_{3;3}
+\frac{14}{3}L_{2;1111}\right.\notag\\
&~~~~~~~~~~~~~~~~
\left.+\frac{5}{2}S_6+6S_{4;2}+\frac{1}{2}S_{3;3}+\frac{1}{2}S_{2;1111}\right)+\cdots
\end{align}
 \label{Fk}
\end{subequations}
where the ellipsis stand for higher mass terms.
By inspecting these formulas, one realizes that the odd instanton contributions are simpler than the
even instanton ones since the latter involve both the ``long" sums $L_{n;m_1\cdots}$ and
the ``short" sums $S_{n;m_1\cdots}$. This observation will turn out to be useful in the following.

\section{Prepotential and recursion relations for the $B_r$ and $C_r$ theories}
\label{secn:recursion}

In this section we show that the instanton corrections to the prepotential can be resummed
into (quasi-)modular forms of $\Gamma_0(2)$ along the lines discussed in Section~\ref{secn:sduality}.
To do so we write the quantum prepotential $f$ as an expansion in the hypermultiplet mass $m$, namely
\begin{equation}
\begin{aligned}
f &=f^{\mathrm{1-loop}}+f^{\mathrm{inst}}=\sum_{n=1}  f_n
~,
\end{aligned}
\label{fgn1}
\end{equation}
with  $f_n \sim m^{2n}$.  The perturbative 1-loop term can be compactly written as
(see for example \cite{D'Hoker:1999ft})
\begin{equation}
\label{F1loop}
\begin{aligned}
 f^{\mathrm{1-loop}} & = \frac{1}{4}\sum_{\alpha\in\Psi} \Bigl[
 -(\alpha\cdot a)^2 \log\left(\frac{\alpha\cdot a}{\Lambda}\right)^2 +
 (\alpha\cdot a+ m)^2 \log\left(\frac{\alpha\cdot a+m}{\Lambda}\right)^2
  \Bigr]~,
\end{aligned}
\end{equation}
where $\alpha$ is an element of the root system $\Psi$ of the gauge algebra.
Expanding $f^{\mathrm{1-loop}}$ for small values of $m$,
all odd powers cancel upon summing over positive and negative roots and in the end, neglecting all
$a$-independent terms, we find
\begin{eqnarray}
f^\mathrm{{1-loop}} & =&  \frac{m^2}{4} \sum_{\alpha\in\Psi}  \log\left(\frac{\alpha\cdot a}{\Lambda}\right)^2
- \sum_{n=2}^\infty  \frac{m^{2n}}{4n(n-1)(2n-1)} \left(L_{2n-2}+S_{2n-2}\right)\label{fn1loop}\\
& =& \frac{m^2}{4} \sum_{\alpha\in\Psi}  \log\left(\frac{\alpha\cdot a}{\Lambda}\right)^2 -\frac{m^4}{24} \left(L_2+S_2\right)
-\frac{m^6}{120}\left(L_4+S_4\right) -
\frac{m^8}{336} \left(L_6+S_6\right) - \cdots~.\nonumber
\end{eqnarray}
The instanton part of the prepotential can be determined from the recursion relation (\ref{recursionG})
which for the $B_r$ and $C_r$ series reads
\begin{equation}
\frac{\partial f_n}{\partial E_2}=-\frac{1}{48}\sum_{\ell=1}^{n-1}\, \frac{\partial f_\ell}{\partial a}\cdot
\frac{\partial f_{n-\ell}}{\partial a}
\label{recursion}
\end{equation}
since $n_{\mathfrak{g}}=2$ in these cases.
The starting point of the recursion is $f_1$ which, as we have seen in the previous section,
just receives a contribution at 1-loop:
\begin{equation}
f_1=\frac{m^2}{4} \sum_{\alpha\in\Psi} \log\left(\frac{\alpha\cdot a}{\Lambda}\right)^2~.
\label{f1loop}
\end{equation}
Starting from this, we will recursively determine the \emph{exact} $q$-dependence of the prepotential
order by order in $m^2$. It is important to realize that the recursion relation
(\ref{recursion}) only fixes the $E_2$-dependence of $f_n$ at a given order. The $E_2$-independent contributions will be determined instead by comparing with the perturbative expansion (\ref{fn1loop})
and the microscopic multi-instanton computations described in the previous section.

\subsection{The $C_r$ theories}
\label{subsecn:prepcn}

We begin our analysis from the $\mathcal{N}=2^\star$ $C_r$ theories for which explicit multi-instanton calculations can be performed with relatively little effort up to high values of $k$ as we
have seen in Section~\ref{secn:multiCr}.
The results up to 4-instantons are given in (\ref{Fk}).
Collecting the various powers of $m^{2n}$ and adding the 1-loop contribution (\ref{fn1loop}), we
can rewrite the prepotential coefficients in the following suggestive form
\begin{subequations}
\begin{align}
f_1&=\frac{m^2}{4} \sum_{\alpha\in\Psi}  \log\left(\frac{\alpha\cdot a}{\Lambda}\right)^2~,
\label{f1q}\\
f_2&=-\frac{m^4}{24}\big(1-24q-72 q^2-96q^3-168q^4+\cdots\big)L_2\notag\\
&~~~\,-\frac{m^4}{24}\big(1-24 q^2-72q^4+\cdots\big)S_2~,\label{f2}\\
f_3&=-\frac{m^6}{120}\big(1+720q^2+3840q^3+10800q^4+\cdots\big)L_4\notag\\
&~~~\,+\frac{m^6}{2}\big(q+6q^2+12q^3+28q^4+\cdots\big)L_{2;11}\notag\\
&~~~\,-\frac{m^6}{120}\big(1 + 720 q^4+\cdots\big)S_4+\frac{m^6}{2} \big(q^2 + 6 q^4+\cdots\big)S_{2;11}~,\label{f3}\\
f_4&=-\frac{m^8}{336}\big(1-720q^2-26880q^3-183960q^4+\cdots\big)L_6\notag\\
&~~~\,+m^8\big(6q^2+48q^3+180q^4+\cdots\big)L_{4;2}
+m^8\big(q^2+8q^3+34q^4+\cdots\big)L_{3;3}\notag\\
&~~~\,+\frac{m^8}{24}\big(q+12q^2+36q^3+112q^4+\cdots\big)L_{2;1111}\notag\\
&~~~\,-\frac{m^8}{336}\big(1 - 840 q^4+\cdots\big)S_6+m^8\big(6q^4+\cdots\big)S_{4;2}
+\frac{m^8}{2} \big( q^4+\cdots\big)S_{3;3}\notag\\
&~~~\,+\frac{m^8}{24}\big(q^2+12q^4+\cdots\big)S_{2;1111}~.\label{f4}
\end{align}
\end{subequations}
As discussed in Section~\ref{secn:sduality}, S-duality requires that the $q$-dependent functions in front of the various sums organize into quasi-modular forms of $\Gamma_0(2)$ which is the modular group for the
$C_r$ theories.
In Tab.~4 we display a basis for such modular forms up to weight 12.
\begin{table}[ht]
\begin{center}
\begin{tabular}{|c|c|}
\hline
Weight & Modular forms of $\Gamma_0(2)$   \\
\hline \hline
 2 & $H_2$\phantom{\Big|}\\
\hline
4& $H_2^2$\,,\,$E_4$ \phantom{\Big|}\\
\hline
6 & $H_2^3$\,,\,$E_6$ \phantom{\Big|}
\\
\hline
8 & $H_2^4$\,,\,$E_4 H_2^2$\,,\,$E_4^2$ \phantom{\Big|}
\\
\hline
10 & $H_2^5$\,,\,$E_4^2 H_2$\,,\,$E_4 E_6$\phantom{\Big|}
\\
\hline
12 & $H_2^6$\,,\,$E_4 H_2^4$\,,\,$E_4^3$\,,\,$E_6^2$\phantom{\Big|}
\\
\hline
\end{tabular}
\end{center}
\caption{A basis of modular forms of $\Gamma_0(2)$ up
to weight 12. The number $n_w$ of modular forms of weight $w$ can be obtained by expanding the
generating function $\frac{1+x^2+x^4}{(1-x^4)(1-x^6)}=\sum_w n_w\,x^w= 1+x^2+2x^4+2x^6+3x^8+3x^{10}+4x^{12}+\cdots$.}
\end{table}
There $E_4$ and $E_6$ are the usual Eisenstein series (see also Appendix~\ref{secn:modular}), while $H_2$ is
the modular form of weight 2 defined in (\ref{h}). Combining them with the second Eisenstein series $E_2$, one finds two quasi-modular forms of weight 2, namely $\{ H_2, E_2 \}$,  four quasi-modular forms
of weight 4, namely $\{ H_2^2, E_4, E_2^2, E_2 H_2 \}$,  and so on.
In principle one can use the $q$-expansion of these forms and  fit the 1-loop and
instanton results (\ref{Fk}) for the first few $n$'s.
For instance, using
\begin{equation}
\begin{aligned}
E_2(\tau)& =1-24q-72 q^2 - 96 q^3 - 168 q^4+\cdots~,
\label{e2q}
\end{aligned}
\end{equation}
and comparing with (\ref{f2}), one finds
\begin{equation}
f_2= -\frac{m^4}{24}E_2(\tau)\,L_2-\frac{m^4}{24}E_2(2\tau)\,S_2~.
\end{equation}
Similarly, from (\ref{f3}), (\ref{e2q}) and
\begin{equation}
E_4(\tau)=1-240 q+2160 q^2 +6720  q^3 +17520 q^4+\cdots~,
\label{e4q}
\end{equation}
one gets
\begin{equation}
\begin{aligned}
f_3&= -\frac{m^6}{720}\big(5E_2^2(\tau)+E_4(\tau)\big) L_4
-\frac{m^6}{576}\big(E_2^2(\tau)-E_4(\tau)\big) L_{2;11}\\
&\quad-\frac{m^6}{720}\big(5E_2^2(2\tau)+E_4(2\tau)\big) S_4
-\frac{m^6}{576}\big(E_2^2(2\tau)-E_4(2\tau)\big)S_{2;11}~.
\end{aligned}
\label{f3fin0}
\end{equation}
This method can be used to find the higher prepotential coefficients $f_n$ even if, when $n$ increases, the number of quasi-modular forms that become available increases as well, thus
requiring more and more microscopic multi-instanton data to fix all coefficients. It is therefore much more efficient to exploit the modular anomaly equation (\ref{recursion}).
For $f_2$ we have
\begin{equation}
\label{recn2}
\frac{\partial f_2}{\partial E_2}
= -\frac{1}{48} \frac{\partial f_1}{\partial a}\cdot \frac{\partial f_1}{\partial a}
=-\frac{m^4}{96 }\,  \sum_{\alpha,\beta\in\Psi}
\frac{(\alpha\cdot \beta)}{(\alpha \cdot a)(\beta\cdot a)}
=-\frac{m^4}{48}\,  (2 L_2+ S_2)
\end{equation}
where in the last step we used the fact that only the terms with $\alpha=\pm \beta$ contribute to the sum since
$L_{1;1}=S_{1;1}=0$%
\footnote{The identities for the ``long" and ``short" sums can be proven with methods similar to those
discussed in Appendix D of \cite{Billo}}.  
Integrating over $E_2$, we find
 \begin{equation}
f_2= -\frac{m^4}{24}E_2(\tau)\,L_2-\frac{m^4}{48}\big(E_2(\tau)+H_2(\tau)\big)S_2
\label{f2fin}
\end{equation}
where the $E_2$-independent term has been added in order to match the 1-loop and the 1-instanton contributions in (\ref{f2}). Notice that only the 1-instanton result is used for this; thus the perfect
matching of the higher order coefficients in the $q$-expansion of (\ref{f2fin}) with the explicit
multi-instanton results (\ref{f2}) has to be regarded as a very strong and highly non-trivial consistency check.
Furthermore, using the duplication formulas of the Eisenstein series given in (\ref{duplication}),
one can easily check that the two expressions for $f_2$ given in (\ref{f3fin0}) and (\ref{f2fin}) coincide.
Proceeding in a similar way for the $m^6$ terms, we get%
\footnote{Here, we have simplified the notation by writing $E_2$ and $H_2$ in place of $E_2(\tau)$ and $H_2(\tau)$.}
 \begin{equation}
\label{recn3}
\begin{aligned}
\frac{\partial f_3}{\partial E_2} &= -\frac{1}{24} \frac{\partial f_1}{\partial a}\cdot \frac{\partial f_2}{\partial a}\\
&=-\frac{ m^6}{1152}\,E_2\,(16L_4+4L_{2;11} +4S_4+S_{2;11})
-\frac{m^6}{1152}\,H_2\,(4 S_4+S_{2;11}) ~.
\end{aligned}
\end{equation}
Integrating over $E_2$ and matching the perturbative and the first instanton terms
against (\ref{f3}) leads to
\begin{eqnarray}
f_3&\!=\!& -\frac{m^6}{720}\big(5E_2^2+E_4\big)L_4-\frac{m^6}{576}\big(E_2^2-E_4\big)L_{2;11}\label{f3fin2}\\
&&-\frac{m^6}{2880}\big(5E_2^2+10 E_2 H_2+10H_2^2-E_4\big)S_4
-\frac{m^6}{2304}\big(E_2^2+2 E_2 H_2-4H_2^2+E_4\big)S_{2;11}\nonumber
\end{eqnarray}
 Using the duplication formula (\ref{duplication}), one can show that
 (\ref{f3fin0}) and (\ref{f3fin2}) are the same.
We stress again that this result is \emph{exact} in $\tau$ and that by expanding it
in powers of $q$ one can obtain the contributions at \emph{any} instanton number
and check that they perfectly agree with those computed with the multi-instanton calculus.

Using the recursion relation we have determined also $f_4$. We now collect our results on the prepotential
coefficients for the $C_r$ theories:
\begin{subequations}
\begin{align}
 f_1&=\frac{m^2}{4} \sum_{\alpha\in\Psi}  \log\left(\frac{\alpha\cdot a}{\Lambda}\right)^2~,\label{f1finsp} \\
f_2& = -\frac{m^4}{24}E_2\,L_2-\frac{m^4}{48}\big(E_2+H_2\big)S_2~, \label{f2finsp}\\
f_3&=-\frac{m^6}{720}\big(5E_2^2+E_4\big)L_4-\frac{m^6}{576}\big(E_2^2-E_4\big)L_{2;11} \label{f3finsp}\\
&~~~-\frac{m^6}{2880}\big(5E_2^2+10 E_2 H_2+10H_2^2-E_4\big)S_4
-\frac{m^6}{2304}\big(E_2^2+2 E_2 H_2-4H_2^2+E_4\big)S_{2;11}~, \nonumber \\
f_4&=-\frac{m^8}{90720}\big(175 E_2^3 + 84 E_2 E_4 + 11 E_6\big)L_6 +\frac{m^8}{8640}\big(5 E_2^3 - 3 E_2 E_4 - 2 E_6\big)\big(L_{4;2}+\ft{1}{6}\,L_{3;3}\big) \nonumber\\
&~~~-\frac{m^8}{41472}\big(E_2^3 - 3 E_2 E_4 + 2 E_6)L_{2;1111}\nonumber\\
&~~~-\frac{m^8}{725760}\big(175 E_2^3 + 525 E_2^2 H_2 + 945 E_2 H_2^2 - 84 E_2 E_4
+ 39 E_6+ 560 H_2^3\big)S_6\nonumber\\
&~~~+\frac{m^8}{69120}\big(5 E_2^3 +15 E_2^2 H_2+ 3 E_2 E_4 - 3 E_6 - 20 H_2^3\big)
\big(S_{4,;2} +\ft{1}{12}\,S_{3; 3} + \ft{2}{3}\, L_{3; 3}\big)\nonumber \\
&~~~-\frac{m^8}{331776}\big(E_2^3 + 3 E_2^2 H_2 - 12 E_2 H_2^2+ 3 E_2 E_4 + E_6  + 4 H_2^3\big)
S_{2;1111}~.\label{f4finsp}
\end{align}
\label{fnfinsp}
\end{subequations}
It is interesting to observe that the combinations of Eisenstein series appearing in $f_2$, $f_3$ and in the first two lines of $f_4$ in front
of the ``long" sums are exactly the same that appear also in the prepotential coefficients
$f_2$, $f_3$ and $f_4$ of the ADE theories studied in \cite{Billo}.
Moreover, using the duplication formulas (\ref{duplication}), one can show that the combinations of
Eisenstein series of the last three lines of $f_4$ in (\ref{f4finsp}) become identical to the ones
of the first two lines, but evaluated in $2\tau$ instead of $\tau$, similarly to what happens in $f_2$ and $f_3$.

Finally we observe that for $r=1$, {\it{i.e.}} for $C_1=sp(2)$, there are only two long roots, $\pm2$, and
thus only the sums of the type $L_n$ are non zero. In this case the quantum prepotential $f$
drastically simplifies and reduces to
\begin{equation}
\begin{aligned}
f^{(C_1)}&={m^2}
\log\Big(\frac{2a}{\Lambda}\Big)-\frac{m^4}{48\,a^2}E_2-\frac{m^6}{5760\,a^4}\big(5 E_2^2 + E_4\big)\\
&~~~-\frac{m^8}{2903040\,a^6}\big(175 E_2^3 + 84 E_2 E_4 + 11 E_6\big)+\cdots~.
\end{aligned}
\label{sp2}
\end{equation}
This exactly coincides with the prepotential of the SU(2) $\mathcal{N}=2^\star$ theory \cite{Billo},
as it should be since Sp(2) $\simeq$ SU(2).

\subsubsection*{1-instanton}

Given the previous results,
it is possible to write a very compact expression for the 1-instanton contribution to the prepotential. Indeed, the only terms which have a 1-instanton part are those proportional to $L_{2;11\cdots}$
(as is clear also from (\ref{Fk1})), and one finds
\begin{align}
F_{k=1}&=m^4\sum_{\ell=0} \frac{m^{2\ell}}{\ell!}\,L_{2;{\underbrace{\mbox{\scriptsize{11\ldots1}}}_{\mbox{\scriptsize{$\ell$}}}}}~
=\sum_{\alpha\in\Psi_{\mathrm{L}}}\frac{m^4}{(\alpha\cdot a)^2}\sum_{\ell=0} \frac{m^{2\ell}}{\ell!}
\!\!\!
\sum_{\beta_1\not=\cdots\not=\beta_\ell\in\Psi(\alpha)} \frac{1}{(\beta_1\cdot a)
\cdots(\beta_\ell\cdot a)}\,\notag\\
&=\sum_{\alpha\in\Psi_{\mathrm{L}}}\frac{m^4}{(\alpha\cdot a)^2}\prod_{\beta\in\Psi(\alpha)}\left(1+
\frac{m}{\beta\cdot a}\right)~.\label{F1}
\end{align}
where the intermediate step follows from the definition (\ref{sumsLS}) of the ``long" sums $L_{2;11\cdots}$.
The number of factors in the product above
is given by the dimension of $\Psi(\alpha)$ which, when $\alpha$ is
a long root of $C_r$, is $2r-2$ (see Appendix~\ref{secn:roots}).
Thus, in (\ref{F1})
the highest power of the mass is $m^{2r+2}$. This is precisely the only term which survives in the
decoupling limit
\begin{equation}
q\to 0~~\mbox{and}~~m\to\infty~~~\mbox{with}~~m^{2r+2}q=\widehat\Lambda^{2r+2}~~\mbox{fixed}~,
\label{decoupling}
\end{equation}
when the $N=2^\star$ theory reduces to the pure $N=2$ SYM theory%
\footnote{Note that the exponent $(2r+2)$ is the 1-loop $\beta$-function coefficient for the pure $\mathcal{N}=2$ SYM theory with gauge group Sp($2r$).}.
In this case the 1-instanton prepotential is
\begin{equation}
q\,F_{k=1}\Big |_{\mathcal{N}=2}=\widehat\Lambda^{2r+2}
 \sum_{\alpha\in\Psi_{\mathrm{L}}}\frac{1}{(\alpha\cdot a)^2}
\prod_{\beta\in\Psi(\alpha)}\frac{1}{\beta\cdot a}~.
\label{F1pure}
\end{equation}
This expression perfectly coincides with the known results present in the literature (see for example
\cite{D'Hoker:1996mu,Ennes:1999fb,Shadchin:2004yx,Marino:2004cn} and in particular \cite{Keller:2011ek}), while (\ref{F1}) is
the generalization thereof to the $\mathcal{N}=2^\star$ symplectic theories.

\subsection{The $B_r$ theories}
\label{subsecn:prepbn}
The $\mathcal{N}=2^\star$ theories with non-simply laced orthogonal gauge groups can be treated
exactly as described above. Indeed, the modular group is again $\Gamma_0(2)$, and
the recursive relation, the 1-loop and the 1-instanton microscopic data have exactly the same form 
as in the $C_r$ theories and differ only in the explicit expression 
for the roots. Thus, the results for the $B_r$ models become very similar to those of the symplectic ones 
with the only differences arising from the different relations among the root sums. 
Skipping the intermediate steps, the prepotential for the so($2r+1$) theory turns out to be
\begin{equation}
F_{so(2r+1)}=2\pi\ii\tau\,a^2+ \sum_{n= 1}^\infty  f_n
\label{prep1so}
\end{equation}
where the first few $f_n$'s are
\begin{subequations}
\begin{align}
f_1&=\frac{m^2}{4} \sum_{\alpha\in\Psi}  \log\left(\frac{\alpha\cdot a}{\Lambda}\right)^2~,
\label{f1finso}\\
f_2&= -\frac{m^4}{24}E_2\,L_2-\frac{m^4}{48}\big(E_2+H_2\big)S_2~,
\label{f2finso}\\
f_3&= -\frac{m^6}{720}\big(5E_2^2+E_4\big)L_4-\frac{m^6}{576}\big(E_2^2-E_4\big)L_{2;11}\label{f3finso}\\
&~~~-\frac{m^6}{2880}\big(5E_2^2+10 E_2 H_2+10H_2^2-E_4\big)S_4
-\frac{m^6}{2304}\big(E_2^2+2 E_2 H_2-4H_2^2+E_4\big)S_{2;11}~, \notag\\
f_4&=-\frac{m^8}{90720}\big(175 E_2^3 + 84 E_2 E_4 + 11 E_6\big)L_6\notag\\
&~~~+\frac{m^8}{8640}\big(5 E_2^3 - 3 E_2 E_4 - 2 E_6\big)\Big(L_{4;2}+\frac{1}{12}L_{3;3}
+\frac{1}{96}S_{3;3}\Big)\notag\\
&~~~-\frac{m^8}{41472}\big(E_2^3 - 3 E_2 E_4 + 2 E_6)L_{2;1111}\notag\\
&~~~-\frac{m^8}{725760}\big(175 E_2^3 + 525 E_2^2 H_2 + 945 E_2 H_2^2 - 84 E_2 E_4
+ 39 E_6+ 560 H^3\big)S_6\label{f4finso}\\
&~~~+\frac{m^8}{69120}\big(5 E_2^3 +15 E_2^2 H_2+ 3 E_2 E_4 - 3 E_6 - 20 H_2^3\big)
\Big(S_{4,;2} +\frac{1}{6}S_{3; 3} \Big) \notag\\
&~~~-\frac{m^8}{331776}\big(E_2^3 + 3 E_2^2 H_2 - 12 E_2 H_2^2+ 3 E_2 E_4 + E_6  + 4 H_2^3\big)\label{f4finso}
S_{2;1111}~.\notag
\end{align}
\label{fnfinso}
\end{subequations}
By expanding the above expressions in powers of $q$, one can obtain the multi-instanton contributions
to the prepotential. At the 1-instanton level one finds exactly the same expression
(\ref{Fk1}) or (\ref{F1}), simply with the roots of $C_r$ replaced by those of $B_r$. 
At the 2-instanton level one finds complete agreement with the expressions obtained in 
Section~\ref{secn:multiBr}. For higher $k$ the above formulas can be used to efficiently 
determine the higher instanton contributions to the prepotential, which instead are 
technically difficult to compute with the localization methods. 
Notice that the expressions in (\ref{fnfinso}) are similar but not identical to those
of (\ref{fnfinsp}) since there are a few differences in $f_4$.

For $B_1$, {\it{i.e.}} for so(3), there are just two short roots, $\pm\sqrt{2}$, so that only the sums of the
type $S_n$ are non-vanishing. In this case the quantum prepotential becomes
\begin{eqnarray}
f^{(B_1)}&=&{m^2}
\log\Big(\frac{\sqrt{2}a}{\Lambda}\Big)-\frac{m^4}{48\,a^2}\big(E_2+H_2\big)
-\frac{m^6}{5760\,a^4}\big(5E_2^2+10 E_2 H_2+10H_2^2-E_4\big)\nonumber\\
&&-\frac{m^8}{2903040\,a^8}\big(175 E_2^3 + 525 E_2^2 H_2 + 945 E_2 H_2^2 - 84 E_2 E_4
+ 39 E_6+ 560 H_2^3\big)+\cdots\nonumber
\end{eqnarray}
which, after using the duplication formulas (\ref{duplication}), takes the form
\begin{equation}
\begin{aligned}
f^{(B_1)}&={m^2}
\log\Big(\frac{\sqrt{2}a}{\Lambda}\Big)-\frac{m^4}{24\,a^2}E_2(2\tau)-\frac{m^6}{1440\,a^4}
\Big(5 E_2^2(2\tau) + E_4(2\tau)\Big)\\
&~~-\frac{m^8}{362880\,a^6}\Big(175 E_2^3(2\tau) + 84 E_2(2\tau) E_4(2\tau) + 11 E_6(2\tau)
\Big)+\cdots~.
\end{aligned}
\label{so3}
\end{equation}
The fact that only modular forms evaluated at $2\tau$ appear means that $f^{(B_1)}$ admits an expansion in powers of $q^2$; in other words only the even instantons contribute.
The prepotential (\ref{so3}) perfectly matches that of the SU(2) theory (see also (\ref{sp2})) provided
\begin{equation}
a~\to~\sqrt{2}a\quad\mbox{and}\quad q_{\mathrm{SO}(3)}^2 = q_{\mathrm{SU}(2)}~.
\label{so3su2}
\end{equation}
The above identification of the coupling constants, which practically amounts to replace $2\tau$ with $\tau$
in (\ref{so3}),  is consistent with the fact that SU(2) is the double cover of
SO(3), {\it{i.e.}} $\mathrm{SU}(2)/\mathbb{Z}_2\simeq \mathrm{SO}(3)$.

\subsection{Relations between the $B_r$ and $C_r$ theories}

The similarity of the results (\ref{fnfinso}) and (\ref{fnfinsp}) for the prepotential of the $B_r$ and $C_r$ theories
is not surprising since, as discussed in Section~\ref{secn:sduality},
they are related by a strong/weak-coupling S-duality. Here we check explicitly this relation
exploiting the properties of the quasi-modular forms and of the ``long" and ``short" sums.

The first observation is that $H_2$, $E_2$, $E_4$ and $E_6$ have simple properties under
\begin{equation}
\tau~\to~-\frac{1}{2\tau}~;
\label{Sduality}
\end{equation}
indeed, one can check (see Appendix~\ref{secn:modular}) that%
\footnote{To simplify the notation, here and in the following, when we write $H_2$, $E_2$, $E_4$ and $E_6$,
we mean that these are evaluated at $\tau$.}
\begin{subequations}
\begin{align}
\frac{1}{2\tau^2}\,H_2\big(-\ft{1}{2\tau}\big)&=-H_2~,\label{Hdual}\\
\frac{1}{2\tau^2}\,E_2\big(-\ft{1}{2\tau}\big)&=E_2+H_2+\frac{6}{\pi\ii\tau}~,\label{E2dual}\\
\frac{1}{4\tau^4}\,E_4\big(-\ft{1}{2\tau}\big)&=-E_4+5H_2^2~,\label{E4dual}\\
\frac{1}{8\tau^6}\,E_6\big(-\ft{1}{2\tau}\big)&=E_6+7H_2^3~.\label{E6dual}
\end{align}
\label{duals}
\end{subequations}
We recall again that the transformation (\ref{Sduality}) does not belong to
the modular group $\Gamma_0(2)$ but it is a generator of the S-duality group, which is a discrete
subgroup of $\mathrm{Sl}(2,\mathbb{R})$.

The second observation is that the root systems of $B_r$ and $C_r$ can be mapped into each other by exchanging (and suitably rescaling) long and short roots. As a consequence of this fact,
the ``long" and ``short" sums in the two theories are related in the following way
(see (\ref{LSa})):
\begin{equation}
\begin{aligned}
L_{n;m_1\cdots m_\ell}^{(B_r)}&= \Big(\frac{1}{\sqrt{2}}\Big)^{n+m_1+\cdots+m_\ell}
\,S_{n;m_1\cdots m_\ell}^{(C_r)} ~,\\
S_{n;m_1\cdots m_\ell}^{(B_r)}&= \big(\sqrt{2}\big)^{n+m_1+\cdots+m_\ell}
\,L_{n;m_1\cdots m_\ell}^{(C_r)}~.
\end{aligned}
\label{LS}
\end{equation}
Combining (\ref{duals}) and (\ref{LS}) with the expressions of the prepotential coefficients
in the orthogonal and symplectic theories, we can check how they are non-perturbatively related with
each other. In order to display these relations in a transparent way, we explicitly indicate the dependence on the coupling constant $\tau$, on the vacuum expectation values $a$ (through the root lattice sums)
and on the quasi-modular form $E_2$ by writing $f_n^{(B_r)}(\tau,a,E_2(\tau))$ and $f_n^{(C_r)}(\tau,a,E_2(\tau))$. Then, from (\ref{f2finso}) and (\ref{f2finsp}), it is not difficult
to show that
\begin{eqnarray}
 f_2^{(B_r)}\big(\!\!-\!\ft{1}{2\tau}, \sqrt{2}\tau a,E_2(-\ft{1}{2\tau})\big)&\!\!\!=\!\!& \!
\frac{1}{2\tau^2}\!\!\left[-\frac{m^4}{24}E_2(-\ft{1}{2\tau})L_2^{(B_r)}\!-\!\frac{m^4}{48}\Big(E_2(-\ft{1}{2\tau})+H_2(-\ft{1}{2\tau})\Big)S_2^{(B_r)}\right]
\nonumber\\
\!&=&\!-\frac{m^4}{48}\big(E_2+H_2+\delta\big)S_2^{(C_r)}-\frac{m^4}{24}\big(E_2+\delta\big)L_2^{(C_r)}\nonumber\\
\!&=&\!f_2^{(C_r)}\big(\tau,a,E_2+\delta\big)\phantom{\Bigg|}
\label{f2dual}
\end{eqnarray}
where  $\delta=\frac{6}{\pi\ii\tau}$ as before. More generally one can prove that
\begin{equation}
f_n^{(B_r)}\big(\!-\!\ft{1}{2\tau}, \sqrt{2}\tau a,E_2(-\ft{1}{2\tau})\big)= f_n^{(C_r)}\big(\tau,a,E_2+\delta\big)~.
\label{fng11}
\end{equation}
This is precisely the type of duality relation discussed in Section~\ref{secn:sduality} (see in particular
(\ref{fng1})). Using it together with the recurrence relations, we can therefore verify that
\begin{equation}
\cS\big[F^{(B_r)}\big] =\mathcal{L}\big[F^{(C_r)}\big]
\label{SL2}
\end{equation}
as we anticipated. Of course, also the reverse relation
\begin{equation}
\cS\big[F^{(C_r)}\big] =\mathcal{L}\big[F^{(B_r)}\big]
\label{SL3}
\end{equation}
is true. These relations, which are an extension of the ones described in \cite{Billo} for the ADE theories provide a highly non-trivial test of the S-duality.

\section{Prepotential and recursion relations for the $G_2$ and $F_4$ theories}
\label{secn:G2F4}

In this section we consider $\mathcal{N}=2^\star$ theories with $G_2$ and $F_4$ gauge algebras. Differently from the classical algebras $B_r$ and $C_r$, the ADHM construction of the instanton moduli space is not known for the exceptional algebras and thus in these cases one cannot rely on the localization
techniques to obtain explicit
multi-instanton results. Nevertheless remarkable progress can be made using the methods described
in the previous section. Let us start our discussion from the $G_2$ theory.

\subsection{The $G_2$ theory}
\label{subsecn:G2}

The prepotential for the $G_2$ theory can be written as
\begin{equation}
F=3\pi\ii\tau\,a^2+\sum_{n\geq 1} f_n
\label{FG2}
\end{equation}
with the first term describing the classical part and $f_n$ the mass expansion coefficients of the quantum prepotential. Now the modular group is $\Gamma_0(3)$ \cite{Dorey:1996hx,Argyres:2006qr,Kapustin:2006pk} and we assume as before that $f_n$ are quasi-modular forms of this group.
For $\Gamma_0(3)$ the ring of quasi modular forms is generated by
\begin{equation}
\begin{aligned}
    \{ E_2, K_2, E_4, E_6 \}
\label{eee}
\end{aligned}
\end{equation}
where $K_2$, the modular form of weight 2 defined in (\ref{h}).
By building monomials of these basic elements one can construct a basis for the modular forms of
$\Gamma_0(3)$ with higher weights, as indicated in Tab.~5 up to weight 12.
\begin{table}[ht]
\begin{center}
\begin{tabular}{|c|c|}
\hline
Weight & Modular forms of $\Gamma_0(3)$   \\
\hline \hline
 2 & $K_2$\phantom{\Big|}\\
\hline
4& $K_2^2$\,,\,$E_4$ \phantom{\Big|}\\
\hline
6 & $K_2^3$\,,\,$E_4 K_2$\,,\,$E_6$ \phantom{\Big|}
\\
\hline
8 & $K_2^4$\,,\,$E_4 K_2^2$\,,\,$E_4^2$ \phantom{\Big|}
\\
\hline
10 & $K_2^5$\,,\,$E_4^2 K_2$\,,\,$E_4K_2^3$\,,\,$E_4 E_6$\phantom{\Big|}
\\
\hline
12 & $K_2^6$\,,\,$E_4^2 K_2^2$\,,\,$E_4 K_2^4$\,,\,$E_4^3$\,,\,$E_6^2$\phantom{\Big|}
\\
\hline
\end{tabular}
\end{center}
\caption{A basis of modular forms for the congruence subgroup $\Gamma_0(3)$ of the modular group up
to weight 12. The number $n_w$ of modular forms of weight $w$ can be obtained by expanding the
generating function $\frac{1+x^2+x^4+x^6}{(1-x^4)(1-x^6)}=\sum_w n_w\,x^w= 1+x^2+2x^4+3x^6+3x^8+4x^{10}+5x^{12}+\cdots$.}
\end{table}
Comparing with the table of the modular forms of $\Gamma_0(2)$, we see many similarities but also some differences. For example at weight 6 we now have three independent modular forms instead of two.

Finally, there is another well-known feature of $G_2$ \cite{Goddard:1976qe} that will play a crucial r\^ole in the following,
namely the fact that by exchanging (and suitably rescaling) long and short roots we obtain an equivalent root system with the two axes interchanged. Thus, under this operation the $G_2$ theory is mapped into another theory
with the same symmetry but with the two vacuum expectation values $a_1$ and $a_2$
interchanged. We call this ``dual" algebra $G'_2$.

The prepotential for the $G_2$ theory  will be derived again from the recursion relation and
the $E_2$-independent terms will be
determined by the perturbative 1-loop prepotential $f^{\mathrm{1-loop}}$ and the one-instanton result
\begin{equation}
\begin{aligned}
F^{(G_2)}_{k=1}&=\sum_{\alpha\in\Psi_{\mathrm{L}}}\frac{m^4}{(\alpha\cdot a)^2}\prod_{\beta\in\Psi(\alpha)}\left(1+
\frac{m}{\beta\cdot a}\right)
=m^4 \,L_2+\frac{m^6}{2}\,L_{2;11}+\frac{m^8}{24}\,L_{2;1111}
\end{aligned}
\label{F1g2}
\end{equation}
where the last step follows from the fact that there are only four factors in the product since
$\mathrm{dim}\big( \Psi(\alpha_{\mathrm{L}})\big)=4$ for $G_2$ (see Appendix~\ref{secn:roots}).
Formula (\ref{F1g2}) was checked to be valid both for theories with ADE in \cite{Billo} and BC gauge groups
in the previous section, and will be assumed to be valid also for $G_2$.
This assumption is well justified not only by the fact that the 1-instanton contribution takes this form in
all other groups considered so far, but also by the fact that the last term in (\ref{F1g2}), which is the one surviving in the pure $\mathcal{N}=2$ $G_2$ theory
where the adjoint hypermultiplet is decoupled, exactly matches the 1-instanton result obtained long-ago in
\cite{Ito:1997bd} from the Picard-Fuchs approach to the Seiberg-Witten curve of the $G_2$ theory,
and more recently in \cite{Keller:2011ek} from coherent states of W-algebras.

The recursion relation for $G_2$ reads
\begin{equation}
\frac{\partial f_n}{\partial E_2}=-\frac{1}{72}\sum_{\ell=1}^{n-1}\, \frac{\partial f_\ell}{\partial a}\cdot
\frac{\partial f_{n-\ell}}{\partial a}
\label{recursiong2fin}
\end{equation}
Again the starting point is $f_1$ which has only the 1-loop part
\begin{equation}
f_1=\frac{m^2}{4} \sum_{\alpha\in\Psi}  \log\left(\frac{\alpha\cdot a}{\Lambda}\right)^2~,
\label{f1G2}
\end{equation}
 On the other hand, according to our working hypothesis, $f_2$ must be proportional to
$L_2$ and $S_2$, since it has to be homogeneous of degree $-2$ in the vacuum expectation values, with
coefficients that are linear combinations of $E_2$ and $K_2$, since it has to be a quasi-modular form
of $\Gamma_0(3)$ with weight 2. Therefore, matching the 1-loop and 1-instanton terms we get
\begin{equation}
\begin{aligned}
f_2&=-\frac{m^4}{24}\,E_2(\tau)\,L_2 - \frac{m^4}{72}\big(E_2(\tau) + 2K_2(\tau)\big)S_2\\
&=-\frac{m^4}{24}\,E_2(\tau)\,L_2-\frac{m^4}{24}\,E_2(3\tau)\,S_2
\end{aligned}
\label{f2G2}
\end{equation}
where the last step is a consequence of the triplication formula
\begin{equation}
E_2(3\tau) = \frac{1}{3}\big(E_2(\tau)+2 K_2(\tau)\big)~.
\label{e23tau}
\end{equation}
The same result follows by solving the recursion relation (\ref{recursiong2fin})
\begin{equation}
\frac{\partial f_2}{\partial E_2}=-\frac{1}{72}\frac{\partial f_1}{\partial a}\cdot
\frac{\partial f_{1}}{\partial a}
\label{recursiong2}
\end{equation}
and matching with the perturbative and 1-instanton terms.
By expanding $f_2$ in powers of $q$ we can obtain all multi-instanton contributions to the prepotential
that are proportional to $m^4$. As we already remarked, for $G_2$ there is no known ADHM construction
of the instanton moduli space and no explicit multi-instanton calculations can be performed.
Thus, our result represents a prediction for the higher instanton terms.

Applying the same method at order $m^6$, it is quite straightforward to get
\begin{eqnarray}
f_3&\!=\!& -\frac{m^6}{720}\big(5E_2^2+E_4\big)L_4-\frac{m^6}{576}\big(E_2^2-E_4\big)L_{2;11}\label{f3g2}\\
&&-\frac{m^6}{6480}\big(5E_2^2+20 E_2 K_2+30K_2^2-E_4\big)S_4
-\frac{m^6}{5184}\big(E_2^2+4 E_2 K_2-6K_2^2+E_4\big)S_{2;11}\nonumber
\end{eqnarray}
where we suppressed the $\tau$ dependence in the right hand side to simplify the notation.
It is interesting to observe that using the triplication formulas (\ref{e23tau}) the quasi-modular forms appearing
in front of $S_4$ and $S_{2;11}$ can be written exactly like the ones appearing in front of $L_4$ and
$L_{2;11}$, respectively, but evaluated in $3\tau$ instead of $\tau$.
The same triplication formulas allow us to show that under
\begin{equation}
\tau~\to~-\frac{1}{3\tau}~,
\label{Sduality3}
\end{equation}
the Eisenstein series and $K_2$ have simple transformation properties, namely
\begin{subequations}
\begin{align}
\frac{1}{3\tau^2}\,K_2\big(-\ft{1}{3\tau}\big)&=-K_2~,\label{Kdual}\\
\frac{1}{3\tau^2}\,E_2\big(-\ft{1}{3\tau}\big)&=E_2+2K_2+\frac{6}{\pi\ii\tau}~,\label{E2dual3}\\
\frac{1}{9\tau^4}\,E_4\big(-\ft{1}{3\tau}\big)&=-E_4+10K_2^2~,\label{E4dual3}\\
\frac{1}{27\tau^6}\,E_6\big(-\ft{1}{3\tau}\big)&=-E_6-7E_4 K_2+35K_2^3~.\label{E6dual3}
\end{align}
\label{duals3}
\end{subequations}
Furthermore, from the properties of the $G_2$ root lattice, it follows that
\begin{equation}
\begin{aligned}
L_{n;m_1\cdots m_\ell}^{(G_2)}&= \Big(\frac{1}{\sqrt{3}}\Big)^{n+m_1+\cdots+m_\ell}
\,S_{n;m_1\cdots m_\ell}^{(G'_2)} ~,\\
S_{n;m_1\cdots m_\ell}^{(G_2)}&= \big(\sqrt{3}\big)^{n+m_1+\cdots+m_\ell}
\,L_{n;m_1\cdots m_\ell}^{(G'_2)}
\end{aligned}
\label{LSg2}
\end{equation}
where $G_2$ and $G'_2$ are dual to each other \cite{Goddard:1976qe} as discussed above.
Combining (\ref{LSg2}) with the transformation rules (\ref{duals3}), one can check that
both $f_2$ and $f_3$ satisfy the expected duality relations
\begin{equation}
f_n^{(G_2)}\big(-\ft{1}{3\tau},\sqrt{3}\tau a,E_2(-\ft{1}{3\tau})\big)=f_n^{(G'_2)}\big(\tau,a,E_2+\delta\big)~.
\label{fndual3}
\end{equation}
Using the recursion formula (\ref{recursiong2fin}), imposing the duality relations (\ref{fndual3}) and matching
with the 1-loop and the 1-instanton results, we managed to determine also the $m^8$-terms of the
prepotential. We now collect all our findings for the $G_2$ theory up to $m^8$:
\begin{subequations}
\begin{align}
f_1 &= \frac{m^2}{4} \sum_{\alpha\in\Psi}  \log\left(\frac{\alpha\cdot a}{\Lambda}\right)^2 ~,\label{f1g2fin}\\
f_2&=-\frac{m^4}{24}\,E_2\,L_2 - \frac{m^4}{72}\big(E_2 + 2K_2\big)S_2~,\label{f2g2fin}\\
f_3&\!=\! -\frac{m^6}{720}\big(5E_2^2+E_4\big)L_4-\frac{m^6}{576}\big(E_2^2-E_4\big)L_{2;11}\label{f3g2fin}\\
&-\frac{m^6}{6480}\big(5E_2^2+20 E_2 K_2+30K_2^2-E_4\big)S_4
-\frac{m^6}{5184}\big(E_2^2+4 E_2 K_2-6K_2^2+E_4\big)S_{2;11}~,\nonumber\\
f_4&=-\frac{m^8}{90720}\big(175 E_2^3 + 84 E_2 E_4 + 11 E_6\big)L_6+\frac{m^8}{8640}\big(5 E_2^3 - 3 E_2 E_4 - 2 E_6\big)L_{4;2}\label{f4g2fin}\\
&-\frac{m^8}{186624}\big(52E_2^3 + 81 E_2^2 K_2- 42 E_2 E_4 +
162 E_2 K_2^2+44 E_6+ 189 E_4 K_2  - 486 K_2^3)L_{2;1111}\nonumber\\
&-\frac{m^8}{2949440}\big(175E_2^3 +\!1050 E_2^2 K_2-84 E_2 E_4 +\!
2940E_2 K_2^2-\!11 E_6-\!245 E_4 K_2 +\!3465 K_2^3\big)S_6\nonumber\\
&+\frac{m^8}{233280}
\big(5 E_2^3 +30 E_2^2K_2 + 3 E_2 E_4 +30 E_2 K_2^2+2 E_6 +20 E_4 K_2+90 K_2^3\big) S_{4,;2}\nonumber\\
&-\frac{m^8}{5038848}\big(52 E_2^3 +231 E_2^2K_2 + 42 E_2 E_4 +42 E_2 K_2^2-44 E_6 -35 E_4 K_2-288 K_2^3\big)
S_{2;1111}\nonumber~.
\end{align}
\end{subequations}
We remark that the $E_2$-dependence in $f_4$ is completely determined
by the recursion relation (\ref{recursiong2fin}) which, however, can not fix the
combinations of the three independent modular forms of $\Gamma_0(3)$ with weight 6, namely $E_6$, $E_4 K_2$ and $K_2^3$, that appear in front of the various sums. To fix these combinations some extra
information is therefore needed beside the 1-loop and 1-instanton data. In the absence of explicit multi-instanton results, we have used the duality relations (\ref{fndual3}). These
impose very severe restrictions on the coefficients, and the fact that all resulting constraints are
mutually compatible, thus allowing for a solution, is a very strong consistency check on our approach.

The previous formulas and their properties allow us to conclude that
\begin{equation}
\cS\big[F^{(G_2)}\big] =\mathcal{L}\big[F^{(G'_2)}\big]~.
\label{SLG2}
\end{equation}
Of course also the reverse relation is true.

\subsection{The $F_4$ theory}
\label{subsecn:F4}
The $\mathcal{N}=2^\star$ theory with gauge group $F_4$ can be analysed in exactly the same way as discussed above. 
In this case the modular group is $\Gamma_0(2)$, like in the $B_r$ and $C_r$ theories.
Again the recursive relation, the 1-loop and the 1-instanton microscopic data match those of $B_r$ and 
$C_r$ models and we have checked that the prepotential coefficients up to $f_3$ are given by exactly the 
same formulas (\ref{f1finsp}), (\ref{f2finsp}) and (\ref{f3finsp}), 
with the ``long" and ``short" sums written with the roots of $F_4$, and that 
they satisfy the recursion relation (\ref{recursion}). These properties imply
\begin{equation}
\cS\big[F^{(F_4)}\big] =\mathcal{L}\big[F^{(F'_4)}\big]
\label{SLF4}
\end{equation}
and its reverse.

\section{Conclusions}
\label{sec:concl}
In this paper we have extended the results of \cite{Billo} to $\cN=2^\star$ theories with non simply-laced gauge algebras and in particular have studied their behaviour under S-duality.
This extension is far from being trivial. {From} the analysis in $\mathcal{N}=4$ theories one expects
that S-duality be a symmetry for ADE gauge groups. But this cannot be the case for the non-simply laced theories, since S-duality exchanges the $B_r$ and $C_r$ algebras and maps $G_2$ and $F_4$ into the rotated $G'_2$ and $F'_4$ algebras.
All these features and constraints can be put into a consistent picture which allows the
computation of the effective prepotential starting from a modular anomaly equation.
The payoff of this approach is a conjecture for the prepotential of the $G_2$ and $F_4$ gauge theories
for which a direct check is possible only at the 1-instanton level, since the microscopic multi-instanton computations are not available for the exceptional algebras.
At the same time given the number of consistency checks which are passed by our proposal, we are
confident about its validity.

It is amusing to observe that the modular anomaly equation by itself is not enough to completely determine the prepotential coefficients in the mass expansion which in general
need the perturbative 1-loop and some microscopic instanton computations in order to be fixed.
Nonetheless even if the latter are not available for the $G_2$ and $F_4$ cases our proposal stands
at the same level of the computations for the classical orthogonal or symplectic algebras
given that our results are written in a form that makes them independent of the details of the various root systems. One final remark: the results presented here overlook the contribution of the $\Omega$-background.
Such contributions could be easily incorporated by the means of the same methods employed in our companion paper \cite{Billo} for the ADE theories. However, in order not to make the formulas too involved, we have preferred to omit them from our presentation.

\vskip 1.5cm
\noindent {\large {\bf Acknowledgments}}
\vskip 0.2cm
We thank Carlo Angelantonj, Sujay Ashok, Massimo Bianchi, Eleonora Dell'Aquila and Igor Pesando for discussions.

The work of M.B., M.F. and A.L. is partially supported  by the Compagnia di San Paolo
contract ``MAST: Modern Applications of String Theory'' TO-Call3-2012-0088.

\vskip 1cm

\appendix
\section{Notations and conventions for the root systems}
\label{secn:roots}
In this appendix we list our conventions for the root systems of the non-simply laced groups. We consider
both the classical algebras $B_r=\mathrm{so}(2r+1)$ and $C_r= \mathrm{sp}(2r)$, and
the two exceptional ones $F_4$ and $G_2$.
In all cases, we denote by $\Psi$, $\Psi_{\mathrm{L}}$ and $\Psi_{\mathrm{S}}$, respectively,
the set of all roots $\alpha$, the set of the long roots $\alpha_{\mathrm{L}}$ and the set of
the short roots $\alpha_{\mathrm{S}}$. Of course one has $\Psi=\Psi_{\mathrm{L}}\cup \Psi_{\mathrm{S}}$.

Given a root $\alpha\in\Psi$, we define its corresponding co-root $\alpha^{\!\vee}$ as
\begin{equation}
\alpha^{\!\vee}= \frac{2}{(\alpha\cdot\alpha)}\,\alpha~,
\label{coroot}
\end{equation}
and introduce the two sets
\begin{equation}
\begin{aligned}
\Psi(\alpha)&=\left\{\beta\in\Psi \, : \,\alpha^{\!\vee}\cdot\beta=1\right\}~,\\
\Psi^{\vee}(\alpha)&=\left\{\beta\in\Psi \, : \,\alpha\cdot\beta^{\vee}=1\right\}~.
\end{aligned}
\end{equation}

\subsection*{The roots of $B_r$}
\label{subsecn:bn}
Let $\{\mathbf{e}_i\,;\,1\le i\le r\}$ be the standard orthonormal basis in $\mathbb{R}^r$.
The set $\Psi_{\mathrm{L}}$ of the long roots of $B_r$ is
\begin{equation}
\Psi_{\mathrm{L}}=\left\{\pm \sqrt{2}\,\mathbf{e}_i\pm\sqrt{2}\, \mathbf{e}_j\,;\,1\le i<j\le r\right\}~,
\label{longBn}
\end{equation}
with all possible signs,
while the set $\Psi_{\mathrm{S}}$ of the short roots is
\begin{equation}
\Psi_{\mathrm{S}}=\left\{\pm \sqrt{2}\,\mathbf{e}_i\,;\,1\le i\le r\right\}~.
\label{shortBn}
\end{equation}
Any long root $\alpha_\mathrm{L}\in \Psi_{\mathrm{L}}$
has length $2$, while any short root $\alpha_\mathrm{S}\in \Psi_{\mathrm{S}}$ has length $\sqrt{2}$. Therefore we have $n_{\mathfrak{g}}=2$.
It is easy to see that
\begin{equation}
\mathrm{ord}\big(\Psi_{\mathrm{L}}\big)= 2r^2-2r~,\quad
\mathrm{ord}\big(\Psi_{\mathrm{S}}\big)= 2r~,
\label{dimensBn}
\end{equation}
so that the total number of roots is $2r^2$ which is indeed the order of $\Psi$.

According to (\ref{coroot}), we have
\begin{equation}
\alpha^{\!\vee}_{\mathrm{L}}=\frac{1}{2}\,\alpha_{\mathrm{L}}~,\quad
\alpha^{\!\vee}_{\mathrm{S}}=\alpha_{\mathrm{S}}~.
\end{equation}
It is not difficult to show that
\begin{equation}
\begin{aligned}
\mathrm{ord}\big( \Psi(\alpha_{\mathrm{L}})\big)&=4r-6= 2h^{\!\vee}-4~,\\
\mathrm{ord}\big( \Psi(\alpha_{\mathrm{S}})\big)&=0= 4 h-4h^{\!\vee}-4~,\\
\mathrm{ord}\big( \Psi^{\vee}(\alpha_{\mathrm{L}})\big)&=4r-8=-2h+4h^{\!\vee}-4~,\\
\mathrm{ord}\big( \Psi^{\vee}(\alpha_{\mathrm{S}})\big)&=2r-2=3h-2h^{\!\vee}-4~,
\end{aligned}
\label{dimBn}
\end{equation}
where $h=2r$ is the Coxeter number and $h^{\!\vee}=2r-1$ is the dual Coxeter number for $B_r$.

\subsection*{The roots of $C_r$}
For the symplectic algebra $C_r$ the long roots are given by
\begin{equation}
\Psi_{\mathrm{L}}=\left\{\pm2 \,\mathbf{e}_i\,;\,1\le i\le r\right\}
\label{longCn}
\end{equation}
where again $\{\mathbf{e}_i\,;\,1\le i\le r\}$ is the standard orthonormal basis in $\mathbb{R}^r$.
The short roots are instead given by
\begin{equation}
\Psi_{\mathrm{S}}=\left\{\pm \mathbf{e}_i\pm \mathbf{e}_j\,;\,1\le i<j\le r\right\}~,
\label{shortCn}
\end{equation}
with all possible signs.
Any long root $\alpha_\mathrm{L}\in \Psi_{\mathrm{L}}$
has length $2$, while any short root $\alpha_\mathrm{S}\in \Psi_{\mathrm{S}}$ has length $\sqrt{2}$, like for the $B_r$ groups. Therefore also in this case we have $n_{\mathfrak{g}}=2$.
Its easy to see that
\begin{equation}
\mathrm{ord}\big(\Psi_{\mathrm{L}}\big)= 2r~,\quad
\mathrm{ord}\big(\Psi_{\mathrm{S}}\big)= 2r^2-2r ~,
\label{dimensCn}
\end{equation}
so that the total number of roots is $2r^2$ which is the order of $\Psi$.

According to (\ref{coroot}), we have
\begin{equation}
\alpha^{\!\vee}_{\mathrm{L}}=\frac{1}{2}\alpha_{\mathrm{L}}~,\quad
\alpha^{\!\vee}_{\mathrm{S}}=\alpha_{\mathrm{S}}~.
\end{equation}
It is not difficult to show that
\begin{equation}
\begin{aligned}
\mathrm{ord}\big( \Psi(\alpha_{\mathrm{L}})\big)&=2r-2= 2h^{\!\vee}-4~,\\
\mathrm{ord}\big( \Psi(\alpha_{\mathrm{S}})\big)&=4r-8= 4 h-4h^{\!\vee}-4~,\\
\mathrm{ord}\big( \Psi^{\vee}(\alpha_{\mathrm{L}})\big)&=0=-2h+4h^{\!\vee}-4~,\\
\mathrm{ord}\big( \Psi^{\vee}(\alpha_{\mathrm{S}})\big)&=4r-6=3h-2h^{\!\vee}-4~,
\end{aligned}
\label{dimCn}
\end{equation}
where $h=2r$ is the Coxeter number and $h^{\!\vee}=r+1$ is the dual Coxeter number for $C_r$.

As is well-known \cite{Goddard:1976qe}, there is a duality between the formulas
for $B_r$ and those for $C_r$ under the exchange of long and short roots, which can be explicitly
verified on the above results.

\subsection*{The roots of $F_4$}
$F_4$ has forty-eight roots; half of them are long roots and half are short roots. Denoting by
$\{\mathbf{e}_1,\mathbf{e}_2,\mathbf{e}_3,\mathbf{e}_4\}$ the standard
orthonormal basis in $\mathbb{R}^4$, the twenty-four long roots are
\begin{equation}
\Psi_{\mathrm{L}}=\left\{\pm \sqrt{2}\,\mathbf{e}_i\pm\sqrt{2}\, \mathbf{e}_j\,;\,1\le i<j\le 4\right\}~,
\label{longF4}
\end{equation}
while the twenty-four short roots are
\begin{equation}
\Psi_{\mathrm{S}}=\left\{\pm \sqrt{2}\,\mathbf{e}_1,\pm \sqrt{2}\,\mathbf{e}_2,\pm \sqrt{2}\,\mathbf{e}_3,\pm \sqrt{2}\,\mathbf{e}_4,\frac{1}{\sqrt{2}}\big(\pm\mathbf{e}_1\pm\mathbf{e}_2
\pm\mathbf{e}_3\pm\mathbf{e}_4\big)\right\}~.
\label{shortF4}
\end{equation}
The long roots have length $2$, while the short roots have length $\sqrt{2}$. Thus, according to (\ref{coroot}), we have
\begin{equation}
\alpha^{\!\vee}_{\mathrm{L}}=\frac{1}{2}\alpha_{\mathrm{L}}~,\quad
\alpha^{\!\vee}_{\mathrm{S}}=\alpha_{\mathrm{S}}~;
\end{equation}
moreover, $n_{\mathfrak{g}}=2$.

Given the explicit expressions (\ref{longF4}) and (\ref{shortF4}), it is easy to see that
\begin{equation}
\begin{aligned}
\mathrm{ord}\big( \Psi(\alpha_{\mathrm{L}})\big)&=14~,\quad
\mathrm{ord}\big( \Psi(\alpha_{\mathrm{S}})\big)=8~,\\
\mathrm{ord}\big( \Psi^{\vee}(\alpha_{\mathrm{L}})\big)&=8~,\quad
\mathrm{ord}\big( \Psi^{\vee}(\alpha_{\mathrm{S}})\big)=14~,
\end{aligned}
\label{dimF4}
\end{equation}

It is interesting to observe that by exchanging (and suitably rescaling) long and short roots, we obtain an equivalent root system; the algebra associated to this dual root system is called $F'_4$.

\subsection*{The roots of $G_2$}
For $G_2$ the six long roots are given by
\begin{equation}
\Psi_{\mathrm{L}}=\left\{\pm \frac{3}{\sqrt{2}}\,\mathbf{e}_1\pm\sqrt{\frac{3}{2}}\,\mathbf{e}_2\,,\,\pm\sqrt{6}\,\mathbf{e}_2\right\}~,
\label{longG2}
\end{equation}
while the six short roots are given by
\begin{equation}
\Psi_{\mathrm{S}}=\left\{\pm\sqrt{2}\,\mathbf{e}_1\,,\,
\pm \frac{1}{\sqrt{2}}\,\mathbf{e}_1\pm\sqrt{\frac{3}{2}}\,\mathbf{e}_2\right\}
\label{shortG2}
\end{equation}
where $\mathbf{e}_1$ and $\mathbf{e}_2$ form the standard orthonormal basis in $\mathbb{R}^2$.
With this choice , the long roots have length $\sqrt{6}$, while the short roots have length $\sqrt{2}$. Thus, according to (\ref{coroot}), we have
\begin{equation}
\alpha^{\!\vee}_{\mathrm{L}}=\frac{1}{3}\alpha_{\mathrm{L}}~,\quad
\alpha^{\!\vee}_{\mathrm{S}}=\alpha_{\mathrm{S}}~;
\end{equation}
moreover, $n_{\mathfrak{g}}=3$.

Given the explicit expressions (\ref{longG2}) and (\ref{shortG2}), it is easy to see that
\begin{equation}
\begin{aligned}
\mathrm{ord}\big( \Psi(\alpha_{\mathrm{L}})\big)&=4~,\quad
\mathrm{ord}\big( \Psi(\alpha_{\mathrm{S}})\big)=2~,\\
\mathrm{ord}\big( \Psi^{\vee}(\alpha_{\mathrm{L}})\big)&=2~,\quad
\mathrm{ord}\big( \Psi^{\vee}(\alpha_{\mathrm{S}})\big)=4~,
\end{aligned}
\label{dimG2}
\end{equation}

It is interesting to observe that by exchanging (and suitably rescaling) long and short roots, we obtain an equivalent root system in which $\mathbf{e}_1$ and $\mathbf{e}_2$ are exchanged. We call the algebra
associated to this dual root system $G'_2$.

We summarize in the following table the properties for the various algebras that are useful for the calculations presented in the main text:
\begin{table}[ht]
\begin{center}
\begin{tabular}{|c||c|c|c|c|c|c|c|}
\hline
$\phantom{\Big|}$
&{\small{$\mathrm{dim}$}}
&{\small{$\mathrm{rank}$}}
&$h^{\!\vee}$
&{\small{$\mathrm{ord}\big(\Psi_{\mathrm{L}}\big)$}}
&{\small{$\mathrm{ord}\big(\Psi_{\mathrm{S}}\big)$}}
&{\small{$\mathrm{ord}\big( \Psi(\alpha_{\mathrm{L}})\big)$}}
&{\small{$\mathrm{ord}\big( \Psi^{\vee}(\alpha_{\mathrm{S}})\big)$}}
\\
\hline
\hline
$B_r\phantom{\Big|}\!\!$
&$r(2r+1)$
&$r$
&$2r-1$
&$2r^2-2r$
&$2r$
&$4r-6$
&$2r-2$\\
\hline
$C_r\phantom{\Big|}\!\!$
&$r(2r+1)$
&$r$
&$r+1$
&$2r$
&$2r^2-2r$
&$2r-2$
&$4r-6$\\
\hline
$F_4\phantom{\Big|}\!\!$
&$52$
&$4$
&$9$
&$24$
&$24$
&$14$
&$14$\\
\hline
$G_2\phantom{\Big|}\!\!$
&$14$
&$2$
&$4$
&$6$
&$6$
&$4$
&$4$\\
\hline
\end{tabular}
\end{center}
\label{tab:mod_spec}
\caption{The main properties of the non-simply laced algebras.}
\end{table}

\section{Modular forms}
\label{secn:modular}

In this appendix we collect the main formulas for the modular functions used in the main text.

\paragraph{$\bullet$} We adopt the standard definitions for the Jacobi $\theta$-functions:
\begin{equation}
 \theta\sp{a}{b}(v|\tau)=\sum_{n\in \Z} \ee^{\pi\ii\tau(n-\frac{a}{2})^2}
\, \ee^{2\pi \ii (n-\frac{a}{2})(v-\frac{b}{2})}~.
\label{th}
\end{equation}
We also define the $\theta$-constants:
$\theta_2(\tau)\equiv\theta\sp{1}{0}(0|\tau)$,
$\theta_3(\tau)\equiv\theta\sp{0}{0}$ and $\theta_4\equiv\theta\sp{0}{1}(0|\tau)$. Their $q$-expansions
are
\begin{equation}
\begin{aligned}
\theta_2(\tau)&=2q^{\frac{1}{8}}\big(1 + q + q^3 + q^{6}+\cdots\big)~,\\
\theta_3(\tau)&=1 + 2 q^{\frac{1}{2}} + 2 q^2 + 2 q^{\frac{9}{2}} + 2 q^{8}+\cdots~,\\
\theta_4(\tau)&=1 - 2 q^{\frac{1}{2}} + 2 q^2 -2 q^{\frac{9}{2}} + 2 q^{8}+\cdots~,
\end{aligned}
\label{thetaexp}
\end{equation}
where $q=\ee^{2\pi\ii\tau}$.

\paragraph{$\bullet$} The Dedekind $\eta$-function is defined as
\begin{equation}
\eta(\tau)= q^\frac{1}{24} \prod_{n=1}^\infty (1-q^n)~.
\end{equation}

\paragraph{$\bullet$} Under the Sl$(2,\mathbb{Z})$ generators
\begin{equation}
T~:~\tau~\to \tau+1~,~~~S~:~\tau~\to~-\frac{1}{\tau}
\label{STb}
\end{equation}
we have
\begin{equation}
\begin{aligned}
&T~:~~~~ \theta_3(\tau) ~\leftrightarrow~ \theta_4(\tau)
~,~~~ \theta_2(\tau)~ \rightarrow ~\ee^{\frac{\ii\pi}{4}}\, \theta_2(\tau)
~,~~~\eta(\tau)~ \rightarrow ~ \ee^{\frac{\ii\pi}{12}}\,\eta(\tau)~,\\
&S~:~~~~ \frac{\theta_2(\tau)}{\eta(\tau)} ~ \leftrightarrow ~
\frac{\theta_4(\tau)}{\eta(\tau)}~,~~~  \frac{\theta_3(\tau)}{\eta(\tau)}~  \rightarrow ~\frac{\theta_3(\tau)}{\eta(\tau)}
~,~~~ \eta(\tau)~\rightarrow~\sqrt{-\ii\tau}\,\eta(\tau)~.
\end{aligned}
\label{modtran1}
\end{equation}

\paragraph{$\bullet$} The Eisenstein series are defined by
\begin{equation}
\label{E246}
\begin{aligned}
E_2(\tau) & = 1 - 24 \sum_{n=1}^\infty \sigma_{1}(n)\, q^{n}
= 1 - 24 q -72 q^2 - 96 q^3 -168 q^4 +\cdots~,\\
E_4(\tau)& = 1 + 240 \sum_{n=1}^\infty \sigma_{3}(n)\, q^{n}
= 1 + 240 q + 2160 q^2 + 6720 q^3+ 17520 q^4 +\cdots~,\\
E_6(\tau) & = 1 - 504 \sum_{n=1}^\infty \sigma_{5}(n)\, q^{2n}
= 1 - 504 q - 16632 q^2 - 122976 q^3 - 532728 q^4 ~,
\end{aligned}
\end{equation}
where $\sigma_{k}(n)$ is the sum of the $k$-th power of the divisors
of $n$, i.e., $\sigma_k(n) = \sum_{d|n} d^k$. The Eisenstein
series $E_4$ and $E_6$ can be expresses as polynomials in the $\theta$-constants
according to
\begin{equation}
\label{eistotheta}
\begin{aligned}
E_4(\tau) & = \frac 12 \big(\theta_2^8(\tau) + \theta_3^8(\tau) + \theta_4^8(\tau)\big)~,\\
E_6(\tau)& = \frac 12 \big(\theta_3^4(\tau) + \theta_4^4(\tau)\big)
 \big(\theta_2^4(\tau) + \theta_3^4(\tau)\big) \big(\theta_4^4(\tau)- \theta_2^4(\tau)\big)~.
\end{aligned}
\end{equation}
The series $E_2$, $E_4$ and $E_6$
are connected among themselves by logarithmic $q$-derivatives
and form a sort of a ``ring'':
\begin{equation}
\label{Eisring}
\begin{aligned}
q\,\frac{\partial E_2(\tau)}{\partial q}& = \frac{1}{12}\big(E_2^2(\tau) -E_4(\tau) \big)~,\\
q\,\frac{\partial E_4(\tau)}{\partial q}& = \frac{1}{3}\big(E_4(\tau)  E_2(\tau) -E_6(\tau) \big)~,\\
q\,\frac{\partial E_6(\tau)}{\partial q}& = \frac{1}{2}\big(E_6(\tau)  E_2(\tau) -E_4^2(\tau)\big) ~.
\end{aligned}
\end{equation}
The series $E_4$ and $E_6$ are modular forms of $\mathrm{Sl}(2,\mathbb{Z})$ with weight 4 and 6
respectively, while $E_2$ is a quasi-modular form of weight 2 with
an anomalous shift. In particular under
$S$, we have
\begin{equation}
\begin{aligned}
E_2(-\ft{1}{\tau})&=\tau^2\,E_2(\tau)+\frac{6\tau}{\pi\ii}~,\quad
E_4(-\ft{1}{\tau})&=\tau^4\,E_4(\tau)~,\quad
E_6(-\ft{1}{\tau})&=\tau^6\,E_6(\tau)~.
\end{aligned}
\label{eismod}
\end{equation}

\paragraph{$\bullet$} The Eisenstein series satisfy the following duplication formulas
\begin{equation}
\begin{aligned}
E_2(2\tau)&= \frac{1}{2}\big(E_2(\tau)+H_2(\tau)\big)~,\\
E_4(2\tau)&=\frac{1}{4}\big(\!-E_4(\tau)+5H_2^2(\tau)\big)~,\\
E_6(2\tau)&=\frac{1}{8}\big(E_6(\tau)+7H_2^3(\tau)\big)
\end{aligned}
\label{duplication}
\end{equation}
where
\begin{equation}
H_2(\tau)=\frac{1}{2}\big(\theta_3^4(\tau)+\theta_4^4(\tau)\big)=1+24 q + 24 q^2 + 96 q^3 + 24 q^4+144 q^5+\cdots~.
\end{equation}
The function $H_2(\tau)$ is a modular form of weight 2 under the congruence subgroup $\Gamma_0(2)
\subset \mathrm{Sl}(2,\mathbb{Z})$, and is such that
\begin{equation}
H_2(-\ft{1}{2\tau})= -2\tau^2\,H_2(\tau)~.
\label{H2t}
\end{equation}
Combining the modular transformation properties (\ref{eismod}) with the duplication formulas (\ref{duplication}) and also with (\ref{H2t}), one can prove that
\begin{equation}
\begin{aligned}
E_2(-\ft{1}{2\tau}\big)&=2\tau^2\big(E_2(\tau)+H_2(\tau)\big)+\frac{12\tau}{\pi\ii}~,\\
E_4(-\ft{1}{2\tau}\big)&=4\tau^4\big(\!-E_4(\tau)+5H_2^2(\tau)\big)~,\phantom{\big|}\\
E_6(-\ft{1}{2\tau}\big)&=8\tau^6\big(E_6(\tau)+7H_2^3(\tau)\big)~.\phantom{\Big|}
\end{aligned}
\label{eisdual2t}
\end{equation}

\paragraph{$\bullet$} The Eisenstein series satisfy the following triplication formulas
\begin{equation}
\begin{aligned}
E_2(3\tau)&= \frac{1}{3}\big(E_2(\tau)+2 K_2(\tau)\big)~,\\
E_4(3\tau)&=\frac{1}{9}\big(\!-E_4(\tau)+10K_2^2(\tau)\big)~,\\
E_6(3\tau)&=\frac{1}{27}\big(\!-E_6(\tau)-7E_4(\tau)\,K_2(\tau)+35K_2^3(\tau)\big)
\end{aligned}
\label{triplication}
\end{equation}
where
\begin{equation}
K_2(\tau)=\left[\left(\frac{\eta^3(\tau)}{\eta(3\tau)}\right)^3+\left(\frac{3\eta^3(3\tau)}{\eta(\tau)}\right)^3\,\right]^{\frac{2}{3}}
=1 + 12 q + 36 q^2 + 12 q^3 + 84 q^4 + 72 q^5+\cdots
\end{equation}
The function $K_2(\tau)$ is a modular form of weight 2 under the congruence subgroup $\Gamma_0(3)
\subset \mathrm{Sl}(2,\mathbb{Z})$, and is such that
\begin{equation}
K_2(-\ft{1}{3\tau})= -3\tau^2\,K_2(\tau)~.
\label{K3t}
\end{equation}
Combining the modular transformation properties (\ref{eismod}) with the triplication formulas (\ref{triplication}) and also with (\ref{K3t}), one can prove that
\begin{equation}
\begin{aligned}
E_2(-\ft{1}{3\tau}\big)&=3\tau^2\big(E_2(\tau)+2K_2(\tau)\big)+\frac{18\tau}{\pi\ii}~,\\
E_4(-\ft{1}{3\tau}\big)&=9\tau^4\big(\!-E_4(\tau)+10K_2^2(\tau)\big)~,\phantom{\big|}\\
E_6(-\ft{1}{3\tau}\big)&=27\tau^6\big(\!-E_6(\tau)-7E_4(\tau)\,K_2(\tau)+35
K_2^3(\tau)\big)~.\phantom{\Big|}
\end{aligned}
\label{eisdual3t}
\end{equation}

\providecommand{\href}[2]{#2}\begingroup\raggedright\endgroup

\end{document}